\documentclass[sigconf,nonacm]{acmart}

\usepackage[range-phrase={-}, range-units=single]{siunitx}
  \DeclareSIUnit{\sample}{Sa}
\usepackage[acronym]{glossaries}
\newacronym{suit}{SUIT}{Software Updates for Internet of Things}
\newacronym{ota}{OTA}{Over-the-Air}

\newacronym{ake}{AKE}{Authenticated Key Exchange}
\newacronym{crqc}{CRQC}{Cryptographically Relevant Quantum Computers}
\newacronym{dh}{DH}{Diffie-Hellman}
\newacronym{dhke}{DHKE}{Diffie-Hellman Key Exchange}
\newacronym{dsa}{DSA}{Digital Signature Algorithm}
\newacronym{ec}{EC}{Elliptic Curve}
\newacronym{ecdh}{ECDH}{Elliptic Curve Diffie-Hellman}
\newacronym{ecdhe}{ECDHE}{Elliptic Curve Diffie-Hellman Ephemeral}
\newacronym{ecdsa}{ECDSA}{the Elliptic Curve Digital Signature Algorithm}
\newacronym{fors}{FORS}{Forest of Random Subsets}
\newacronym{fts}{FTS}{Few-Time Signature}
\newacronym{hmac}{HMAC}{Hash-based Message Authentication Code}
\newacronym{hss}{HSS}{Hierarchical Signature System}
\newacronym{kem}{KEM}{Key Encapsulation Mechanism}
\newacronym{lwe}{LWE}{Learning With Errors}
\newacronym{ots}{OTS}{One-Time Signature}
\newacronym{pkc}{PKC}{Public-Key Cryptosystem}
\newacronym{pke}{PKE}{Public-Key Encryption}
\newacronym{puf}{PUF}{Physical Unclonable Function}
\newacronym{qc}{QC}{Quantum Computing}
\newacronym{rot}{RoT}{Root of Trust}
\newacronym{rsa}{RSA}{the Rivest–Shamir–Adleman cryptosystem}
\newacronym{sis}{SIS}{Short Integer Solution}
\newacronym{svp}{SVP}{Shortest Vector Problem}
\newacronym{uov}{UOV}{Unbalanced Oil and Vinegar}

\newacronym{ca}{CA}{Certificate Authority}
\newacronym{dn}{DN}{Distinguished Name}
\newacronym{pki}{PKI}{Public Key Infrastructure}
\newacronym{ta}{TA}{Trust Anchor}
\newacronym{api}{API}{Application Programming Interface}
\newacronym{cn}{CN}{Common Name}
\newacronym{csr}{CSR}{Certificate Signing Request}
\newacronym{curl}{cURL}{}
\newacronym{ebnf}{EBNF}{extended Backus-Naur form}
\newacronym{gnutls}{GnuTLS}{}
\newacronym{gtaapi}{GTA-API}{Generic Trust Anchor API}
\newacronym{gui}{GUI}{Graphical User Interface}
\newacronym{jdk}{JDK}{Java Development Kit}
\newacronym{mac}{MAC}{Message Authentication Code}
\newacronym{mitm}{MitM}{Man-in-the-Middle}
\newacronym{openssl}{OpenSSL}{}
\newacronym{os}{OS}{Operating System}
\newacronym{pem}{PEM}{Privacy Enhanced Mail}
\newacronym{ra}{RA}{Registration Authority}
\newacronym{rng}{RNG}{Random Number Generator}
\newacronym{san}{SAN}{Subject Alternative Names}
\newacronym{w3m}{W3M}{}
\newacronym{wget}{Wget}{GNU Wget}
\newacronym{xml}{XML}{Extensible Markup Language}
\newacronym{xsl}{XSL}{Extensible Stylesheet Language}

\newacronym{aisec}{AISEC}{Fraunhofer Institute for Applied and Integrated Security}
\newacronym{anima}{ANIMA}{Autonomic Networking Integrated Model and Approach}
\newacronym{bsi}{BSI}{Bundesamt für Sicherheit in der Informationstechnik}
\newacronym{eu}{EU}{European Union}
\newacronym{iec}{IEC}{International Electrotechnical Commission}
\newacronym{ieee}{IEEE}{Institute of Electrical and Electronics Engineers}
\newacronym{ietf}{IETF}{Internet Engineering Task Force}
\newacronym{isa}{ISA}{International Society of Automation}
\newacronym{nist}{NIST}{National Institute of Standards and Technology}
\newacronym{oqs}{OQS}{Open Quantum Safe}
\newacronym{rfc}{RFC}{Request For Comments}
\newacronym{tcg}{TCG}{Trusted Computing Group}
\newacronym{www}{WWW}{World Wide Web}

\newacronym[plural=DevIDs,firstplural=Device Identifiers (DevIDs)]{devid}{DevID}{Device Identifier}
\newacronym[plural=IDevIDs,firstplural=Initial Device Identifiers (IDevIDs)]{idevid}{IDevID}{Initial Device Identifier}
\newacronym[plural=JRCs,firstplural=Join Registrar and Coordinators (JRCs)]{jrc}{JRC}{Join Registrar and Coordinator}
\newacronym[plural=LDevIDs,firstplural=Locally Significant Device Identifiers (LDevIDs)]{ldevid}{LDevID}{Locally Significant Device Identifier}
\newacronym[plural=MASAs,firstplural=Manufacturer Authorized Signing Authorities (MASAs)]{masa}{MASA}{Manufacturer Authorized Signing Authority}
\newacronym{acp}{ACP}{Autonomic Control Plane}
\newacronym{ani}{ANI}{Autonomic Networking Infrastructure}
\newacronym{jws}{JWS}{JSON Web Signature}
\newacronym{prm}{PRM}{Pledge Responder Mode}

\newacronym[plural=CRs,firstplural=Component Requirements (CRs)]{cr}{CR}{Component Requirement}
\newacronym[plural=HMIs,firstplural=Human-Machine Interfaces (HMIs)]{hmi}{HMI}{Human-Machine Interface}
\newacronym[plural=IACS,firstplural=Industrial Automation and Control Systems (IACS)]{iacs}{IACS}{Industrial Automation and Control System}
\newacronym[plural=ISPs,firstplural=Integration Service Providers (ISPs)]{isp}{ISP}{Integration Service Provider}
\newacronym[plural=MES,firstplural=Manufacturing Execution System (MESs)]{mes}{MES}{Manufacturing Execution System}
\newacronym[plural=MSPs,firstplural=Maintenance Service Providers (MSPs)]{msp}{MSP}{Maintentance Service Provider}
\newacronym[plural=PLCs,firstplural=Programmable Logic Controllers (PLCs)]{plc}{PLC}{Programmable Logic Controller}
\newacronym[plural=REs,firstplural=Requirement Enhancements (REs)]{re}{RE}{Requirement Enhancement}
\newacronym[plural=SL-As,firstplural=Achieved Security Levels (SL-As)]{sl-a}{SL-A}{Achieved Security Level}
\newacronym[plural=SL-Cs,firstplural=Capability Security Levels (SL-Cs)]{sl-c}{SL-C}{Capability Security Level}
\newacronym[plural=SL-Ts,firstplural=Target Security Levels (SL-Ts)]{sl-t}{SL-T}{Target Security Level}
\newacronym[plural=SLs,firstplural=Security Levels (SLs)]{sl}{SL}{Security Level}
\newacronym[plural=SRs,firstplural=System Requirements (SRs)]{sr}{SR}{System Requirement}

\newacronym[plural=ASAs,firstplural=Autonomic Service Agents (ASAs)]{asa}{ASA}{Autonomic Service Agent}
\newacronym[plural=HTAs,firstplural=Hardware Trust Anchors (HTAs)]{hta}{HTA}{Hardware Trust Anchor}
\newacronym[plural=LLNs,firstplural=Low-Power and Lossy Networks (LLNs)]{lln}{LLN}{Low-Power and Lossy Network}
\newacronym[plural=OEMs,firstplural=Original Equipment Manufacturer (OEMs)]{oem}{OEM}{Original Equipment Manufacturer}
\newacronym[plural=OIDs,firstplural=Object Identifiers (OIDs)]{oid}{OID}{Object Identifier}
\newacronym[plural=SEs,firstplural=Secure Elements (SEs)]{se}{SE}{Secure Element}
\newacronym[plural=TPMs,firstplural=Trusted Platform Modules (TPMs)]{tpm}{TPM}{Trusted Platform Module}
\newacronym[plural=VDIs,firstplural=Virtual Disk Images (VDIs)]{vdi}{VDI}{Virtual Disk Image}
\newacronym[plural=VMs,firstplural=Virtual Machines (VMs)]{vm}{VM}{Virtual Machine}
\newacronym{cli}{CLI}{Command-line Interface}
\newacronym{dmz}{DMZ}{Demilitarized Zone}
\newacronym{dos}{DoS}{Denial of Service}
\newacronym{e2e}{E2E}{end-to-end}
\newacronym{ecc}{ECC}{Elliptic Curve Cryptography}
\newacronym{email}{e-mail}{Electronic Mail}
\newacronym{eid}{eID}{electronic Identity Document}
\newacronym{gpio}{GPIO}{General Purpose Input/Output}
\newacronym{iot}{IoT}{Internet of Things}
\newacronym{it}{IT}{Information Technology}
\newacronym{json}{JSON}{JavaScript Object Notation}
\newacronym{maas}{MaaS}{MASA-as-a-Service}
\newacronym{mdm}{MDM}{Mobile Device Management}
\newacronym{nat}{NAT}{Network Address Translation}
\newacronym{opcua}{OPC UA}{Open Platform Communications Unified Architecture}
\newacronym{ot}{OT}{Operational Technology}
\newacronym{scada}{SCADA}{Supervisory Control and Data Acquisition}
\newacronym{spi}{SPI}{Serial Peripheral Interface}
\newacronym{ssd}{SSD}{Solid State Drive}
\newacronym{ssi}{SSI}{Self Sovereign Identity}
\newacronym{tofu}{TOFU}{Trust On First Use}
\newacronym{uri}{URI}{Uniform Resource Identifier}
\newacronym{url}{URL}{Uniform Resource Locator}
\newacronym{var}{VAR}{Value-Added Reseller}
\newacronym{wot}{WoT}{Web of Trust}

\newacronym{cpu}{CPU}{Central Processing Unit}
\newacronym{mcu}{MCU}{Microcontroller Unit}
\newacronym{pc}{PC}{Personal Computer}
\newacronym{ram}{RAM}{Random Access Memory}
\newacronym{rom}{ROM}{Read Only Memory}
\newacronym{uC}{µC}{Microcontroller}
\newacronym{ui}{UI}{User Interface}
\newacronym{uP}{µP}{Microprocessor}
\newacronym{zsbl}{ZSBL}{Zero-Stage Boot Loader}
\newacronym{otp-m}{OTP memory}{One Time Programmable memory}

\newacronym{pi3b}{Pi3B}{Raspberry Pi 3B}
\newacronym{pi3b+}{Pi3B+}{Raspberry Pi 3B+}
\newacronym{pi4}{Pi4}{Raspberry Pi 4}

\newacronym{ble}{BLE}{Bluetooth Low Energy}
\newacronym{brski}{BRSKI}{Bootstrapping Remote Secure Key Infrastructure}
\newacronym{cbrski}{cBRSKI}{Constrained BRSKI}
\newacronym{cmc}{CMC}{Certificate Management Protocol using the Cryptographic Message Syntax}
\newacronym{cmp}{CMP}{Certificate Management Protocol}
\newacronym{cms}{CMS}{Cryptographic Message Syntax}
\newacronym{coap}{CoAP}{Constrained Application Protocol}
\newacronym{coaps}{CoAPS}{Secure Constrained Application Protocol}
\newacronym{crl}{CRL}{Certificate Revocation List}
\newacronym{ct}{CT}{Certificate Transparency}
\newacronym{dhcp}{DHCP}{Dynamic Host Configuration Protocol}
\newacronym{dns}{DNS}{Domain Name System}
\newacronym{dpp}{DPP}{Device Provisioning Protocol}
\newacronym{dtls}{DTLS}{Datagram Transport Layer Security}
\newacronym{est}{EST}{Enrollment over Secure Transport}
\newacronym{ftp}{FTP}{File Transfer Protocol}
\newacronym{grasp}{GRASP}{Generic Autonomic Signaling Protocol}
\newacronym{html}{HTML}{Hypertext Markup Language}
\newacronym{http}{HTTP}{Hypertext Transfer Protocol}
\newacronym{https}{HTTPS}{Hypertext Transfer Protocol Secure}
\newacronym{imap}{IMAP}{Internet Message Access Protocol}
\newacronym{ip}{IP}{Internet Protocol}
\newacronym{ipsec}{IPsec}{Internet Protocol Security}
\newacronym{lan}{LAN}{Local Area Network}
\newacronym{ldap}{LDAP}{Lightweight Directory Access Protocol}
\newacronym{ocsp}{OCSP}{Online Certificate Status Protocol}
\newacronym{plcom}{PLC}{Power Line Communication}
\newacronym{pq}{PQ}{Post-Quantum}
\newacronym{pqc}{PQC}{Post-Quantum Cryptography}
\newacronym{rest}{REST}{Representational State Transfer}
\newacronym{rpl}{RPL}{IPv6 Routing Protocol for Low-Power and Lossy Networks}
\newacronym{scep}{SCEP}{Simple Certificate Enrolment Protocol}
\newacronym{sip}{SIP}{Session Initiation Protocol}
\newacronym{smtp}{SMTP}{Simple Mail Transfer Protocol}
\newacronym{ssh}{SSH}{Secure Shell}
\newacronym{ssl}{SSL}{Secure Sockets Layer}
\newacronym{sztp}{SZTP}{Secure Zero Touch Provisioning}
\newacronym{tcp}{TCP}{Transmission Control Protocol}
\newacronym{tls}{TLS}{Transport Layer Security}
\newacronym{wan}{WAN}{Wide Area Network}
\newacronym{xslt}{XSLT}{Extensible Stylesheet Language Transformation}

\newacronym{bw}{BW}{Bandwidth}
\newacronym{rtt}{RTT}{Round-Trip Time}
\newacronym{ttfb}{TTFB}{Time To First Byte}

\newacronym{ek}{EK}{Endorsement Key}
\newacronym{hse}{HSE}{Hardware Security Engine}
\newacronym{hsm}{HSM}{Hardware Security Module}
\newacronym{iak}{IAK}{Initial Attestation Key}
\newacronym{lak}{LAK}{Local Attestation Key}

\newacronym{ev}{EV}{Electric Vehicle}
\newacronym{ecu}{ECU}{Electronic Control Unit}
\newacronym{cp}{CP}{Charging Point}

\newglossaryentry{802.1ar}
{
    name={IEEE 802.1AR},
    description={An IEEE standard for secure device identifiers and device enrollment in industrial control systems}
}

\newglossaryentry{isaiec}
{
    name={IEC 62443},
    description={An international series of standards that address cybersecurity for operational technology in automation and control systems}
}

\newglossaryentry{pledge}
{
    name={pledge},
    plural={pledges},
    description={An device that wishes to bootstrap via BRSKI into a network}
}
\usepackage{tabularx}
\usepackage{arydshln} % \hdashline: dotted rule 
\usepackage{multirow} 
\usepackage[htt]{hyphenat} % break texttt

\title[PQC in Lightweight Virtualization]{Benchmarking Post-Quantum Cryptography in Lightweight Virtualization Environments on Embedded Hardware}

\author{Nikolai Puch}
\orcid{0009-0000-6259-9846}
\affiliation{%
  \institution{Technical University of Munich}
  \city{Munich}
  \country{Germany}
}
\affiliation{  
  \institution{Fraunhofer AISEC}
  \city{Garching}
  \country{Germany}
}
\email{nikolai.puch@aisec.fraunhofer.de}
\correspondingauthor

\author{Chi Hieu Ta}
\orcid{0009-0008-0072-3627}
\affiliation{%
  \institution{CarByte Engineering GmbH}
  \city{Rülzheim}
  \country{Germany}
}
\email{chi-hieu.ta@carbyte.de}
\email{chihieu.ta@tum.de}

\author{Moritz Beckel}
\orcid{0009-0005-6402-9624}
\affiliation{%
  \institution{Technical University of Munich}
  \city{Munich}
  \country{Germany}
}
\email{moritz.beckel@tum.de}


\begin{document}

\begin{abstract}
% Abstract 
%
\gls{pqc} is being deployed at the same time as embedded systems increasingly adopt lightweight virtualization for workload isolation and security.
Both trends change performance characteristics, yet their interaction is not well understood.
To address this research gap, we present a measurement study of \gls{pqc} primitives on embedded-class ARM hardware under three execution environments with a shared software stack:
native execution, a Docker container, and a Unikraft unikernel running under QEMU.
We benchmark five signature and five key encapsulation mechanism families, and two classical algorithms each, for comparison.
We evaluate them using different parameter sets for a total of around 70~configurations, measuring execution time, memory, and energy per operation.
To better gauge the impact on applications, we evaluated 13 TLS~1.3 cipher combinations as well.

We find that container overhead is negligible for primitive computation, whereas unikernel overhead depends on the algorithm.
For % ECDSA, RSA,
CROSS, FrodoKEM, Classic McEliece, and the NIST-standardized post-quantum families ML-KEM, ML-DSA, and the SHAKE variants of SPHINCS+, the overhead is near-native.
A moderate overhead (\SIrange{1.28}{1.53}{\times}) arises in the alternates BIKE, HQC, and MAYO, and, above all, in Falcon signing (\SIrange{17.8}{19.2}{\times}), while the SHA-2 variants of SPHINCS+ run faster than native (\SIrange{0.54}{0.67}{\times}).
Per-operation energy closely tracks execution time in all environments.
For TLS handshakes, container and unikernel clients need more handshake time and energy per handshake when the cryptographic cost is low, while all three environments converge once expensive post-quantum algorithms dominate the handshake.
In these cases algorithm choice affects per-handshake energy by up to three orders of magnitude, far outweighing the environment.
Overall, virtualization cost is inversely related to cryptographic cost: environment choice matters most for computationally cheap, standardized algorithms, while for expensive schemes, algorithm choice alone dominates performance.

% DSA: Dilithium2/ML-DSA, Falcon, SPHINCS+, MAYO, CROSS
% KEM: BIKE, Classic-McEliece, HQC, Kyber/ML-KEM, FrodoKEM

\end{abstract}

\received{22 August 2026}

\maketitle

\glsresetall

% Introduction. Motivation follows the repository's own framing
% (epqciuoe/Report/document.tex, introduction; 10_paper/sections/01_introduction.tex).
% RQ1/RQ2 are taken verbatim from the old draft (evidence-map.md E21); RQ3 is
% added because power data exists in the repository (E25).
\section{Introduction}
\label{sec:introduction}
A large-scale \gls{crqc} would break the public-key cryptography that secures today's Internet~\cite{Shor1997}.
Even before a \gls{crqc} exists, traffic can be recorded and decrypted retroactively once such machines exist, in so-called Harvest-Now, Decrypt-Later attacks.
Recent advances \cite{Cain2026} and such scenarios have accelerated the transition, with the \gls{eu} aiming for a transition by the end of 2030 for high-risk use cases \cite{EU-Roadmap2025}.
Thus, the migration to \gls{pqc} is already underway: \gls{nist} has standardized ML-KEM~\cite{FIPS203}, ML-DSA~\cite{FIPS204}, and SLH-DSA~\cite{FIPS205}, selected HQC in its fourth round~\cite{Alagic2025}, and announced Falcon for future standardization~\cite{Alagic2022}.
% The TLS relevance argument follows the repo report (document.tex l.75).
\gls{tls} is the most immediate deployment target, as both its key exchange and its authentication must be replaced with \gls{pq} algorithms, at the cost of larger keys and signatures.

At the same time, embedded and edge systems increasingly rely on software-based isolation for easier deployment and security.
Containers are today's dominant mechanism, but they require a full host operating system.
Unikernels aim to become an alternative and promise smaller images, a reduced attack surface, and near-native performance on top of a hypervisor.
To achieve this, unikernels compile an application together with only the OS components it needs~\cite{Madhavapeddy2014}.
% Unikraft in particular reports competitive or better-than-Linux-guest
% performance for common server applications~\cite{Kuenzer2021}.
This is especially interesting for constrained embedded domains, like the automotive domain, where compute, memory and energy are at a premium.

PQC and lightweight virtualization each distinctly impact the system's performance. Their \emph{combination} is largely unexplored: post-quantum algorithms stress memory, entropy sources, and floating-point units in ways classical cryptography does not, and unikernels virtualize exactly these resources differently than containers do.
Operators tasked with choosing between deploying an application utilizing \gls{pqc} in a container or in a unikernel on embedded hardware currently face a lack of benchmark data to guide their decision.
%
% RQs: RQ1/RQ2 verbatim from 10_paper/sections/01_introduction.tex.
This paper addresses this gap and answers the following research questions:
\begin{enumerate}
    \item[\textbf{RQ1}] How do PQC primitives perform in unikernels versus other virtualization solutions?
    \item[\textbf{RQ2}] How does this impact secure communication via \gls{tls}?
    \item[\textbf{RQ3}] What is the energy cost of PQC under the different virtualization options?
\end{enumerate}

To answer them, we build a benchmarking framework that runs an identical cryptographic stack, based on liboqs~0.12.0, natively, in a Docker container, and in a Unikraft unikernel on a Raspberry~Pi~4B, and we measure primitive execution time, memory, and per-operation energy as well as TLS~1.3 handshake throughput and energy.

\paragraph{Contributions}
\begin{itemize}
    \item Automated benchmarking framework covering three execution environments on ARM64 embedded hardware with an identical crypto stack (Section~\ref{sec:setup}).
    \item A primitive-level comparison of 60 \gls{pqc} configurations across the three environments, covering speed, memory, and per-operation energy (Section~\ref{sec:evaluation}).
    \item A TLS~1.3 handshake study over 13~signature/KEM combinations, including energy measurements (Section~\ref{sec:evaluation}).
          % \item An analysis showing that virtualization overhead is inversely           related to cryptographic cost, together with candidate explanations for
          %       algorithm-specific unikernel anomalies
          %       (Section~\ref{sec:discussion}).
\end{itemize}

% Background. Content condensed from 10_paper/sections/20_background.tex and
% epqciuoe/Report/document.tex (background chapter); external claims carry the
% citations that the repository authors already collected. Structure: PQC
% first (what changes), then virtualization (where it runs) -- mirroring the
% two axes of the study.
\section{Background}
\label{sec:background}

\subsection{Post-Quantum Cryptography}

Post-quantum algorithms are commonly grouped by function into \glspl{dsa} and \glspl{kem}, and by the underlying hardness assumption.
\autoref{tab:pqcciphers} gives an overview of the current \gls{nist} algorithms and their status.
% Naming convention: the repo deliberately uses NIST-competition names
% (Kyber/Dilithium) rather than the standardized names (ML-KEM/ML-DSA), see
% 10_paper/sections/20_background.tex l.10. We keep that convention.
Following prior embedded evaluations, we use the NIST-competition names of the algorithms and their liboqs implementations rather than the standardized names, because the implementations we benchmark predate final standard release.

\textbf{Signatures:} 
\emph{Dilithium}, standardized as ML-DSA in FIPS~204~\cite{FIPS204}, is a module-lattice scheme and the primary NIST signature choice.
\emph{Falcon} is a lattice scheme selected for standardization as FIPS~206~\cite{Alagic2022}.
Its current implementations depend on double-precision floating-point arithmetic, which is missing on many embedded platforms. 
However, our evaluation platform provides this.
\emph{SPHINCS+}, standardized as SLH-DSA in FIPS~205~\cite{FIPS205},
is a stateless hash-based scheme with conservative security assumptions but
comparatively expensive signing.
% Additional round-two candidates \cite{Alagic2026} \emph{MAYO} and \emph{CROSS} are examples for an Unbalanced Oil and Vinegar and respectively a  Fiat-Shamir transform based algorithm.
%
\textbf{KEMs:} 
\emph{Kyber} a lattice-based \gls{kem}, standardized as ML-KEM in FIPS~203~\cite{FIPS203}, is the primary \gls{kem} by \gls{nist}. 
\emph{HQC}, a code-based scheme, was selected in NIST's fourth round~\cite{Alagic2025} as an alternative.

% \emph{BIKE} and
% \emph{Classic McEliece} are further code-based candidates, and
% \emph{FrodoKEM} is an unstructured-lattice scheme available with AES- or
% SHAKE-based internal symmetric primitives. ECDHE serves as the classical
% baseline.

\begin{table}[ht!]
    \centering
    \caption{\gls{pqc} algorithms and their publication status.}
    \label{tab:pqcciphers}
    \begin{tabular}{l|ll}
        % \textbf{Cipher}    & \textbf{Type}       & \textbf{Status}              \\
        \multicolumn{3}{l}{\textbf{Hash-based ciphers}}                                      \\ \hline
        % LMS                & stateful \gls{dsa}  & RFC 8554 \cite{RFC8554}                   \\
        % XMSS               & stateful \gls{dsa}  & RFC 8391 \cite{RFC8391}                   \\
        SPHINCS+           & stateless \gls{dsa} & FIPS 205~\cite{FIPS205}              \\
        \multicolumn{3}{l}{\textbf{Lattice-based ciphers}}                                   \\ \hline
        CRYSTALS-Kyber     & \gls{kem}           & FIPS 203~\cite{FIPS203}              \\
        CRYSTALS-Dilithium & \gls{dsa}           & FIPS 204~\cite{FIPS204}              \\
        Falcon             & \gls{dsa}           & FIPS 206 (WiP~\cite{Alagic2026}) \\
        \multicolumn{3}{l}{\textbf{Code-based ciphers}}                                      \\ \hline
        % BIKE               & \gls{kem}           & Round 4 submission           \\
        % Classic McEliece   & \acrshort{pkc}      & Round 4 submission         \\
        HQC                & \acrshort{kem}      & Round 4 selection~\cite{Alagic2025}      \\
    \end{tabular}
\end{table}

\begin{figure}[htb]
    \includegraphics[width=\linewidth]{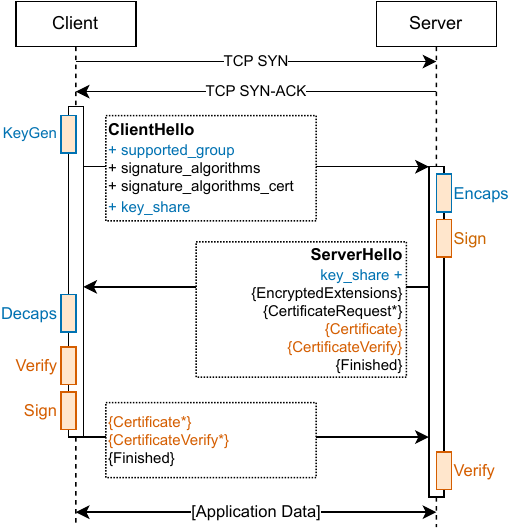}
    \caption{\label{fig:pqtls} Shows the \gls{pq} TLS~1.3 handshake. Orange shows the necessary changes to the \gls{dsa} and blue to the \gls{kem}.}
\end{figure}

\paragraph{PQC in TLS}
Multiple approaches exist for making \gls{tls} quantum resistant \cite{Gonzalez2022}.
The currently favoured method is a \gls{pq} TLS~1.3 handshake, which replaces both the ephemeral key exchange and the authentication via \gls{dsa}.
\glspl{kem} are comparatively fast and small on embedded devices and are only done once per handshake.
The required changes are displayed in \autoref{fig:pqtls}.
However, the \gls{dsa} is applied once per link of the certificate chain, so signature verification cost and certificate size dominate the handshake overhead for many
schemes~\cite{Puch2026,Tasopoulos2022,Gonzalez2022}.

\subsection{Virtualization on Embedded Systems}
% Condensed from 10_paper/sections/20_background.tex §Virtualization; the
% container-vs-hypervisor distinction and the MicroVM exclusion are stated
% there and in the repo report.

\begin{figure}[htb]
    \includegraphics[width=\linewidth]{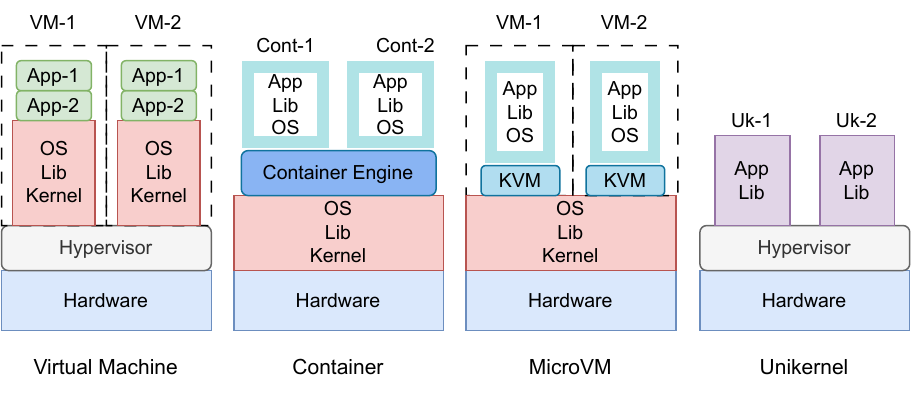}
    \caption{\label{fig:overview-uk} Visual comparison of virtualization architectures}
\end{figure}

Virtualization partitions the resources of a physical machine into isolated instances. Two families are relevant for resource-constrained systems. \emph{OS-level virtualization} (containers, e.g., Docker) isolates processes using kernel features such as namespaces and cgroups; it adds little runtime overhead but shares the host kernel, so its isolation boundary is the kernel interface itself. It is worth mentioning that OS-level virtualization is not a security feature by default and measures are necessary to secure it.
\emph{Hypervisor-based virtualization}, on the other hand, runs guests on a virtual machine monitor. Depending on the use-case and underlying hardware, a complete operating system, or images with smaller footprints, such as unikernels and MicroVMs are virtualized. The latter applies this model to single-purpose workloads instead of providing operating system capabilities. The security property is inherently given by the isolation of the hypervisor between host and guest. A visual comparison between containers, virtual machines, MicroVMs, and unikernels is given in \autoref{fig:overview-uk}. This study omits MicroVMs and compares containers with unikernels.
Extending the comparison to MicroVMs is left to future work.

\paragraph{Unikernels}
A unikernel, also known as library operating system, compiles the application together with only the OS components it uses into a single-address-space image that boots
directly on a hypervisor. One of the advantages, in light of required performance, is the absence of a dedicated user and kernel space. 
With one singular address space no context switches must be performed, thus increasing the overall performance. Ordinarily, removing the separation between user and kernel space would introduce security issues on its own. 
However, with unikernels this is intentional. This issue is resolved by the design of unikernels, as they are compiled and used as single-purpose and read-only binary files. Hence, unikernels are immutable.~\cite{Madhavapeddy2014}
% and no scheduler competition from other processes; system calls become function calls. 

% Entropy note: mechanism verified in epqciuoe/Unikraft/setup.sh (SEED via
% kernel cmdline); relevant later for interpreting results (E15).
Architectural difference matters for cryptographic workloads: in our configuration the unikernel guest has no hardware entropy source of its own. It receives a random seed on the kernel command line at boot (drawn from the host's \texttt{/dev/urandom}) and expands it internally, whereas native and containerized processes obtain entropy from the host kernel (e.g., via \texttt{getrandom}) throughout their lifetime.

% Related work. Anchored on the bibliography curated by the repository authors
% (10_paper/pqc-vs-unikernel.bib, evidence-map.md E19) plus the framing in
% 10_paper/sections/10_relatedwork.tex. Organized in three strands: PQC on
% embedded, PQ-TLS evaluations, unikernel performance. The comparison-gap
% argument (no PQC-in-unikernel measurements) is the paper's positioning; we
% state it as "to the best of our knowledge" since we cannot prove absence.
\section{Related Work}
\label{sec:related}

\paragraph{PQC on embedded systems}
Evaluating post-quantum algorithms on constrained hardware has been an integral part of the NIST standardization process, which included benchmarks on the ARM Cortex-M4 microcontroller~\cite{Alagic2022}.
\texttt{PQClean}~\cite{pqclean.git} collects clean, standalone \textit{C} implementations of these algorithms. In 2026 it was superseded by the \texttt{PQ Code Package}~\cite{pqcode.git}, a collection of open-source implementations maintained within the Linux Foundation.
Targeting \gls{uC} platforms more specifically, \texttt{pqm4}~\cite{pqm4.git} collects optimized implementations of the standardized algorithms or recent standardization candidates for the ARM Cortex-M4.
\texttt{liboqs}~\cite{liboqs.git} integrates many of these implementations and forms the basis for the \texttt{\acrshort{oqs}-provider}~\cite{oqs-provider.git}, which brings \gls{pqc} to OpenSSL~\cite{oqstls}.
% , as well as for the inital \gls{pqc} feature in \texttt{WolfSSL}~\cite{WolfSSL.pqc}.

Based on these libraries, several works benchmark the then-most-recent \gls{pq} algorithms on embedded platforms within specific application domains, including \gls{iot}~\cite{Hanna2025}, industrial control~\cite{Reichert2025}, and automotive systems~\cite{Bos2022,Bos2023}.
Among these domains, \gls{tls} is the most prominently researched application for \gls{pqc} on embedded devices.
Most research in this field likewise builds on \texttt{pqm4} or \texttt{liboqs} and targets the ARM Cortex-M4 and Cortex-A53.
Results show that \gls{dsa} selection and network conditions, owing to the increased key and signature sizes, dominate the performance of \gls{pq} \gls{tls}~\cite{Puch2026}.
Mixed certificate chains~\cite{Schoeffel2022,Paul2022} allow tuning the key/signature size vs. performance tradeoff to the specific use case, while KEMTLS~\cite{Gonzalez2022} offers an alternative approach.
Beyond the Cortex-M4/A53 focus of this body of work, Bürstinghaus-Steinbach et al.~\cite{Burstinghaus-Steinbach2020} benchmarked Kyber and SPHINCS+ inside mbedTLS across a broader range of embedded platforms.
Energy consumption, by contrast, is evaluated far less frequently: Tasopoulos et al.\ measure \gls{pq} TLS~1.3 handshake time on resource-constrained devices~\cite{Tasopoulos2022} and, in follow-up work, its energy consumption~\cite{Tasopoulos2023}.
They find that although Falcon completes its handshake faster, its power consumption exceeds that of Dilithium due to the larger compute share it requires.
The aforementioned mixed-certificate infrastructure can likewise be used to balance such power tradeoffs~\cite{Schoeffel2022}.

\paragraph{Unikernel performance}
Madhavapeddy and Scott were the first to introduce unikernels also known as library operating systems~\cite{Madhavapeddy2014}. Meanwhile, many other instances of unikernel frameworks have been introduced, such as rumprun which is based on the rump kernel, and OSv ~\cite{rumpkernelrumprun_2026, osv_nodate}. In addition to rumprun and OSv, Unikraft is introduced as another framework which reports near-native or better-than-Linux-guest performance for server workloads such as nginx and Redis.
According to the github page of rumprun the last commit was performed six years ago at the time of our work. Further, the work of Kuenzer et al. show that their Unikraft unikernels report a 1.7--2.7$\times$ performance improvement over Linux guests for common server applications, image sizes around 1\,MB, and boot times in the millisecond range. In addition, it outperforms rumprun and OSv in the same use-cases, hence we opted for Unikraft for our tests due to still ongoing development compared to rumprun and better performance compared to rumprun and OSv.~\cite{Kuenzer2021}

% TODO @Chio from Moritz: Ich weis nicht ob das wichtig ist aber mir ist noch aufgefallen in der Unikernel analyse (307) sind die Ergebnisse glaub schon bisschen veraltet, so viel performance kriegt man Unikraft ja nicht mehr 

%Unikraft~\cite{Kuenzer2021} is
%a modular unikernel framework whose authors report a 1.7--2.7$\times$ performance improvement over Linux guests for common server applications, image sizes around 1\,MB, and boot times in the millisecond range. Unikraft targets POSIX/Linux compatibility via a syscall shim layer, which makes it a practical vehicle for porting OpenSSL-based workloads. We use Unikraft~0.18.0 in this study.

\paragraph{Unikernel vs Container}
Unikernel and container have been subjected to comparative benchmarking across various experimental setups. Goethals et al., for instance, compare unikernels against containers in the context of microservices. They conducted their tests on an x86\_64 machine with 4GB of RAM and 160GB HDD which is used as a server, while
client tasks were done on a separate machine to avoid distorting the measurements. While containers were spawned on Ubuntu 18.04, they opted for a XenServer 7.5 for their unikernel setup.
Furthermore, Goethals et al. divided their test cases into single threaded and multithreaded tests. Also, they emphasised that only one container or virtual machine is active during tests.~\cite{Goethals2018}
In~\cite{Plauth2017} a comparison between container and multiple unikernel frameworks is performed. The test cases comprise an HTTP server and key-value store to represent workloads usually found in cloud applications. In this regard, Plauth et al. use an x86\_64 architecture CPU including 2x8GB of RAM, and KVM as their hypervisor for their test environment.
Similar to Plauth et al., Acharya et al. benchmark containers and multiple unikernel frameworks in the context of network functions virtualization. They opt for two server hardware variants to test their application. I.e., one with 8 core x86\_64 CPU and 32GB of DDR4 and another platform with a 64bit ARMv8 and 48 cores with 128GB of DDR4 RAM. Operating system-wise, Acharya et al. use Ubuntu 16.04.3 LTS in conjunction with KVM as hypervisor and QEMU 2.5.0. The performed test cases contain regular CPU performance, as well as memory and network bandwidth benchmarks.~\cite{Acharya2018}
Although not strictly a head-on comparison of two technologies against each other, Bartolomeo et al. propose a hybrid usage of containers and unikernels for edge computing. The goal is to reduce CPU load by orchestrating the usage of either unikernels or containers depending on the use-case in which either performs best. For this, they test both technologies under different circumstances. Bartolomeo et al. target systems involving two server set-ups with x86\_64 and a Raspberry Pi 4 with the ARM Cortex-A72 and 8GB of DDR4 RAM. They use KVM as hypervisor but use an extended version of Oakestra to spawn and orchestrate unikernels and containers in parallel.~\cite{Bartolomeo2025}\\
As we can see, unikernels are mainly used and benchmarked within a server comparable set-up, i.e., using x86\_64 CPUs with a reasonable amount of RAM. While Acharya et al. use a dedicated ARMv8 test environment, it is a server-sized setup, and they specifically stated that unikernel tests on it are excluded due to the lack of stable ARM support of rumprun and OSv~\cite{Acharya2018}. Only Bartolomeo et al. with the Raspberry Pi setup offer a comparable test environment and results to our research, as they use an ARMv8 single board computer and the Unikraft framework to build their unikernels. However, their results still differ in the ultimate goal, as post-quantum algorithms or in general applications involving the usage and performance of encrypted communication is not one of their evaluation criteria. Accordingly, cryptographic workloads, and post-quantum algorithms in particular, have not been part of prior evaluations and research.

% \paragraph{Gap.}
Hence, to the best of our knowledge, no prior work measures \gls{pq} primitives or \gls{pq} \gls{tls} inside lightweight virtualization, neither containers nor unikernels, on embedded hardware, nor compares the environments against each other under an identical cryptographic stack. The present paper bridges this gap.

% Methodology. Grounded in epqciuoe/Benchmark/run_benchmarks.py (stages,
% durations, algorithm lists, TLS combinations), the measurand definitions in
% 10_paper/sections/30_experiment.tex, and analyze_power.py. Design choices
% that are potential threats to validity (server always native; different
% network paths; top-based memory sampling) are stated openly here and
% revisited in the limitations section.
\section{Methodology}
\label{sec:methodology}

Our goal is to isolate the effects of the transition to post-quantum cryptographic on the performance in different execution environments. 
We therefore keep hardware, cryptographic library, libc, and compiler toolchain constant and vary only the environment: \textsc{Native}, \textsc{Container}, and \textsc{unikernel}. 
For these environments we measure three quantities: speed, memory, and power.
We evaluate the isolated primitives alone as well as in the context of a full TLS~1.3 handshake. 
All benchmarks are orchestrated by a single Python driver to make runs reproducible.
% (run_benchmarks.py; stages primitives/tls x speed/power/memory.)

\subsection{Metrics}
\label{subsec:metrics}

\paragraph{Primitive speed}
For each algorithm and operation (key generation, signing, verification for \glspl{dsa} and key generation, encapsulation, decapsulation for \glspl{kem}) we execute the operation in a loop and report the mean time per operation together with its population standard deviation. 
We let each operation run for at least a minimum wall-clock time as well as a minimum number of operations, and then collect the final wall-clock time as well as the number of completed operations. 
The speed evaluation uses wall-clock-time via \texttt{clock\_gettime(CLOCK\_REALTIME)}, while the clock for the timestamps of the power benchmark uses \texttt{perf\_time\_ns}.
The primitive dataset evaluated in this paper uses a \SI{20}{\second} window per operation.
We benchmark 40~signature and 20~\gls{kem} configurations, covering all NIST security levels available in liboqs~0.12.0 for Dilithium, Falcon (including padded variants), SPHINCS+, (SHA-2 and SHAKE, fast and small variants), MAYO, CROSS, BIKE, Classic McEliece, HQC, Kyber, and FrodoKEM (AES and SHAKE variants).

The TLS datasets stem from an earlier measurement campaign, primitive and TLS results are therefore not from the same run.
However, it follows overall a similar approach:
We measure the number of completed TLS~1.3 connections in a 30\,s window using \texttt{openssl s\_time} against \texttt{openssl s\_server}. 
The server always runs natively on the \gls{pi4}, while the client runs in the environment under test.
This choice isolates the client-side handshake cost but implies that each environment reaches the server over a different network path: 
loopback for native, Docker's NAT for the container, and a QEMU bridge network for the unikernel. 
We discuss this in more detail in Section~\ref{sec:limitations}.
% (run_benchmarks.py: run_s_server("native", ...); README commands; E8/E24.)
We report initial (full-handshake) connections.
Certificates are generated per signature algorithm with a single self-signed \gls{ca} and one server certificate.
% (run_benchmarks.py setup_certificates; README PKI commands.)
We evaluate 13~signature/KEM combinations: Dilithium and Falcon paired with Kyber at matching security levels, SPHINCS+-128s with Kyber512, and Dilithium2 paired with BIKE-L1, HQC-128, and FrodoKEM-640 (AES and SHAKE), with RSA-2048+ECDHE and ECDSA+ECDHE as classical baselines. 
The subset follows prior post-quantum TLS studies on embedded hardware~\cite{Tasopoulos2023} to ease cross-comparison.

\paragraph{Memory}
For the primitive campaign the benchmark binary tagged with a special \texttt{pid} flag which is passed to the measurement loop to monitor memory.
The monitor reads \texttt{/proc/\textit{pid}/statm} and \texttt{/proc/\textit{pid}/status} at \SI{200}{\hertz} and \SI{10}{\hertz}, respectively. % persistent file descriptors 
From \texttt{statm} it collects instantaneous virtual-memory (VM) and resident-set-size (RSS) page counts, which are converted to bytes using the system page size.
From \texttt{status} it tracks the kernel-maintained high-water marks \texttt{VmPeak} and \texttt{VmHWM}.
% The monitor terminates when the process exits, detected via \texttt{ESRCH} on a zero-signal \texttt{kill}.
The resulting dataset contains, per operation: mean RSS and mean VM over all \SI{200}{\hertz} samples, the sample-maximum of each, and the kernel-reported \texttt{VmHWM} and \texttt{VmPeak} values.
For the TLS memory stage the driver spawns the client process under the same \texttt{pid} wrapper and then collects memory readings with \texttt{top~-b~-p~\textit{pid}~-d~0.5}.
Mean VIRT and RES values are averaged across all samples.
For the unikernel, the monitored process in both cases is QEMU, so the reported figures reflect the memory footprint of the whole guest as seen by the host, including all guest RAM actually touched. 
This is the deployment-relevant quantity for an operator placing workloads. %, but it is not directly comparable to per-process RSS of the native or container environments.
For the container, the same \texttt{pid} mechanism targets \texttt{containerd-shim-v2}, a host-side management process.
We discuss the resulting comparability implications for the container vs. unikernel comparison in Section~\ref{sec:limitations}.

\paragraph{Power and energy}
The earlier TLS campaign used a FNIRSI FNB58 USB power meter placed inline in the device's USB-C supply. Segment boundaries are detected by current-draw change. The FNB58 runs at \SI[per-mode = symbol]{100}{\sample\per\second} and is logged to a separate device. According to the data sheet the power meter has a resolution of $\SI{0.00001}{volt}$ and $\pm 0.02\%+2$ digits \cite{FNIRSI2026}. Although the manufacturer’s technical specifications claim high static resolution, a formalized error propagation and statistical uncertainty analysis were omitted for this hardware, due to the absence of a certified chain of traceability to primary international metrology standards. Hence, we decided to opt for professional hardware to perform the primitive campaign.
For the primitive campaign, the device is powered by a Rohde\,\&\,Schwarz HMC8043 programmable power supply \cite{man.HMC804x} controlled over SCPI/TCP. The R\&S HMC8043 has a resolution of \SI{1}{\milli\volt} and \SI{0.1}{\milli\ampere}, if $I < \SI{1}{\ampere}$ and logging $< \SI[per-mode = symbol]{100}{\sample\per\second}$. The reading accuracy is $<0.05\% + \SI{2}{\milli\volt}$ and $<0.05\% + \SI{2}{\milli\ampere}$. At \SI[per-mode = symbol]{100}{\sample\per\second} and the resulting resolution of \SI{10}{\milli\second} might cause aliasing, as load changes in the \gls{pi4} can be faster than \SI{50}{\hertz}. However, the long runtime of \SI{20}{\second} ensures that this error is acceptable. The built-in logging function is used to create a trace over the whole campaign.
The trace is then aligned with the per-operation start/stop timestamps recorded by the benchmark driver and alignment is manually verified. This is then used to split the trace per operation. 
We report energy per operation ($E_{op}$), i.e., the integrated segment power divided by the number of iterations executed in the segment.
For TLS we derive energy per completed handshake by dividing the segment energy by the connection count recorded in the same run.
%- 10s laufen lassen, durchschnittliche Power
%- Im Anschluss beliebige Operation
%- power_baseline_mw = über 10s gemittelt -> das von power p_per_iteration abziehen -> ergibt in Tabelle 13 Dilithium Keygen native P erster wert
In terms of power per operation ($P_{op}$) an additional step must be performed to acquire the value. For this, we start our test with the respective environment and leave it in idle state for \SI{10}{\second}. With the R\&S HMC8043 and $< \SI[per-mode = symbol]{100}{\sample\per\second}$ we receive $\approx$ 1000 power samples which in turn is averaged over the time. This power draw serves as the baseline. Subsequently, the operation is started and measured over \SI{20}{\second}. The resulting power is consequently subtracted by the baseline yielding $P_{op}$.

We report this baseline-subtracted value rather than raw power because the idle baseline drifts by up to $\approx$600\,mW between measurement windows independent of which environment is running. Comparatively, this is large in relation to the many operations' own power draw above the idle state which is $\approx$0.7--1.4\,W. So, if not subtracted the raw energy would be dominated by this drift for a substantial share of comparisons.
%
% (Benchmark/hmc804x.py, measure_power.py, analyze_power.py,
%  merge_results.py, HMC804x.md; merged JSON fields e_per_iteration /
%  p_per_iteration; consistency e ≈ P * t verified.)

\paragraph{Formal measurement-error}
We conduct a formal measurement-error estimation according to JCGM GUM-1:2023\cite{GUM.2023}.
Type A is calculated for memory and energy, while type B is only calculated for the energy measurements by the R\&S HMC8043.

% Fehler A
For the type A evaluation we conducted 25 measurement campaigns for the algorithms ML-DSA, ML-KEM, ECDSA, and X25519.
The other parameters were kept the same as in the evaluation runs.
We calculate the standard deviation $s(x)$ over the 25 runs to calculate the uncertainty $u_A(x)$ (see Equation \ref{eq:u-a}).
% Discuss speed not for now, as the way we masure this already includes std dev - one could do the 25 runs additional, but only small benefit
% For the speed the $u_A(speed)= \SIrange{16.33}{13322.45}{\nano\second}$
For the memory we calculate an uncertainty of $u_A(rss) = \SIrange{0.25}{20.89}{\kibi\byte}$, for baseline-subtracted power of $u_A(P) = \SIrange{7.693}{121.267}{\milli\watt}$ and for baseline-subtracted energy of $u_A(E) = \SIrange{0.0008}{5.830}{\milli\joule}$.
%TODO Noch mal schauen, ob sich bei größerem n noch was groß ändert.

% Fehler B
We estimate the type B uncertainty of the energy with equation \ref{eq:uncertainty_general}.
$u_E(E)$ is the uncertainty of the energy measurement, $u_t(E)$ is the uncertainty of the time base, and $u_{jump}(E)$ is the uncertainty of the energy at an edge in the powertrace. 

\begin{equation}
	\label{eq:uncertainty_general}
    u_B(E)      = \sqrt{u_E^2(E) + u_t^2(E) + u_{jump}^2(E)}
\end{equation}

The following values are the basis for the calculation:
From the datasheet of R\&S HMC8043 \cite{man.HMC804x} we obtained the relative error $a_U = a_I = 0.05 \% $, and reading error $b_U = \SI{2}{\milli\volt}$, and $b_I = \SI{2}{\milli\ampere}$ for voltage and current respectively. From measurement of the clock the \gls{pi4} $a_t = \SI{630}{\nano\second}$. 
From the empirical evaluation of our powertrace, we determine: 
    $|\Delta P_s| \cdot \Delta t = \SI{23.371}{\milli\joule}$ and 
    $|\Delta P_e| \cdot \Delta t = \SI{23.893}{\milli\joule}$ for the start and end respectively.
% python .\find_power_jump.py .\results\powertrace_2026-08-10_17-30-08.csv
% Largest rising jump:
%   12:27:50:040 -> 12:27:50:050: 3.0259 W -> 5.3630 W (+2.3371 W)
% Largest falling jump:
%   10:12:51:760 -> 10:12:51:770: 5.9359 W -> 3.5466 W (-2.3893 W)
%
Evaluating equation \ref{eq:uncertainty_voltage_of_energy} and \ref{eq:uncertainty_current_of_energy} over all 537 primitive operations in the power dataset, we calculate the relative uncertainty of voltage and current as a function of energy as $u_U(E)=\SI{51.9}{\milli\joule}$ and $u_I(E)=\SI{189.0}{\milli\joule}$.
Hence, formulas \ref{eq:uncertainty_energy_of_energy}, \ref{eq:uncertainty_time_of_energy}, and \ref{eq:uncertainty_jump} are estimated to be $u_E(E)=\SI{0.19}{joule}$, $u_t(E)=\SI{1.82}{\milli\joule}$, and $u_{jump}(E) = \SI{9.648}{\milli\joule}$. Consequently, equation \ref{eq:uncertainty_general} is evaluated as $u_B(E)=\SI{0.19}{joule}$, $u_{jump}(E)$ is constant across operations and contributes only $5\%$ of the uncertainty of $u_B(E)$ for a typical \SI{20}{\second} segment. $u_t(E)$ is negligible ($\approx 10^{-3}\%$ relative) regardless of operation.

\subsection{Controls}
All binaries are built for ARM64 and statically linked against \texttt{musl 1.2.3}.
The native binary is built inside the same Alpine container as the containerized one and exported, so that \texttt{libc} differences cannot confound the comparison.
% (30_experiment.tex Native paragraph; E6.)
Benchmark processes were bound to a single core with \texttt{os.sched\_setaffinity}, respectively the \texttt{--cpuset-cpus} flag in Docker. 
%
% Device temperature is controlled via \texttt{vcgencmd}.
For the primitives campaign after each operation the temperature is checked, and the campaign is paused until the temperature drops below \SI{50}{\degreeCelsius} to prevent thermal throttling.
The highest momentary peaks we measured was up to \SI{53.0}{\degreeCelsius}, which remains well below the \gls{pi4} default thermal-throttling threshold, so throttling can reasonably be excluded for the current primitive datasets. 
% default thermal-throttling threshold:  $\approx\SI{80.0}{\degreeCelsius}$
% (results/tls_speed.json temp_start/temp_end; merged JSON temp fields; E17.)
%
% Experimental setup. Versions verified from committed build scripts
% (Native/setup.sh, Container/alpine-bench-aarch64, Unikraft/uk-build-aarch64,
% Unikraft/unikraft/version.mk, ssl_uk_bin/.config) and Report/versions.txt
% (E2-E6). QEMU invocation parameters from Unikraft/setup.sh.
\section{Experimental Setup}
\label{sec:setup}

\subsection{Hardware and Host Software}
\begin{figure}[htb]
    \includegraphics[width=\linewidth]{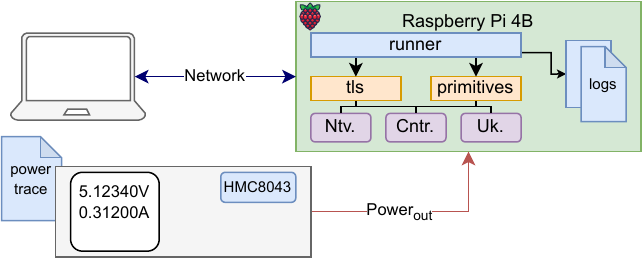}
    \caption{\label{fig:measuresetup} Shows a schematic overview of the measurement setup. Power supply is based on the primitives campaign. In the Raspberry Pi the runner with the benchmark binaries and the different environments is outlined.}
\end{figure}

All experiments are conducted on a Raspberry~Pi~4 Model~B (Broadcom BCM2711, quad-core ARM Cortex-A72, 64-bit). 
The platform was chosen as a widely available microprocessor-class stand-in for automotive or embedded hardware, whose processors are architecturally comparable \cite{nxp-i.MX8}.
% (00_Wissenssammlung/hardware.md: platform selection rationale.)
The Raspberry~Pi~4 has \qty{8}{\giga\byte} of RAM, runs Debian~12 (bookworm) at kernel version \texttt{6.12.34+rpt-rpi-v8} with QEMU~7.2.17, Docker~28.3.1, and
Python~3.11.2 for instrumentation.
% (Report/versions.txt; Report/pi-installed-packages.txt.)
Depending on the campaign the Pi is powered by a Rohde\,\&\,Schwarz HMC8043 programmable power supply or through a FNIRSI FNB58 USB-C meter inline with a power support plug.
For a detailed description see \autoref{subsec:metrics}.
% (Benchmark/measure_power.py IDN string "Rohde&Schwarz,HMC8043,...";
%  Benchmark/HMC804x.md; fnirsi_usb_power_data_logger/.)
\autoref{fig:measuresetup} gives an overview of the measurement setup.

\subsection{Cryptographic Stack}
All three environments use OpenSSL~3.4.1 with liboqs~0.12.0, the native and container variants additionally load oqs-provider~0.8.0 to expose liboqs algorithms through OpenSSL's provider interface, while the unikernel links liboqs statically into the image.
% (Native/setup.sh l.6-8; Container/alpine-bench-aarch64 l.7-9;
%  uk-build-aarch64 l.48-49. Wording per E3 note: oqs-provider is not part of
%  the unikernel build.)
A single benchmark source tree is shared by all three environments.
It is derived from the liboqs speed tests and calls post-quantum algorithms through the native liboqs API. 
We extended the benchmark macro to include a minimum number of iterations, as well as a minimum time, added CSV export for offline analysis, and fixed an unrelated bug in RDTSC.   
We evaluate all \gls{pq} algorithm families supported by liboqs~0.12.0, however, we had to exclude some of the larger parameter sets of CROSS-small and Classic-McEliece. 
As a baseline it also implements the classical ECDSA, ECDHE, X25519, and RSA-2048 through OpenSSL's EVP interface.
For ECDSA and ECDHE the prime256v1 curve is chosen.
% (shared/oqs_speed/: speed_sig.c/speed_kem.c use OQS_SIG/OQS_KEM;
%  speed_vanilla.c uses EVP_PKEY_keygen/EVP_DigestSign etc.)
All binaries are compiled for ARM64 and statically linked against musl~1.2.3.

\subsection{The Three Environments}
\paragraph{Native.}
The baseline runs the benchmark and OpenSSL binaries directly on the host OS.
To keep the libc identical across environments, the native binaries are built inside the same Alpine container used for the container variant and exported to the host.

\paragraph{Container.}
The container variant runs the identical binaries inside a Docker container based on Alpine Linux. 
The TLS client inside the container reaches the server via Docker's host gateway. % (\texttt{host.docker.internal})

\paragraph{Unikernel.}
Unikraft~0.18.0 ("Helene") is used to build unikernels \cite{Kuenzer2021}. Unikraft provides a native port of OpenSSL, however, the version does not support the provider architecture of newer OpenSSL versions. Hence, we compile OpenSSL and liboqs as static libraries into a unikernel using Unikraft. We opted for static library builds, as compared to rewriting the Makefile, this approach yields less configuration overhead. 
Three application images are built: a primitive benchmark, an OpenSSL command-line image restricted to the \texttt{s\_server}, \texttt{s\_client}, and \texttt{s\_time} tools, and a liboqs self-test image used to validate correctness of the port. Porting required patches to musl and to auxiliary-vector handling on ARM64.
% (Unikraft/patches/{musl-aarch64,auxv,getauxval,aligned-alloc}.patch.)
Images run under \texttt{qemu-system-aarch64} with \texttt{-machine virt-cpu max}, a bridged virtual network (172.44.0.0/24), and a 9pfs share for exchanging files with the host. To reduce the overhead for the unikernel we build and sign certificates for the TLS tests on the host system and bootstrap the unikernel via 9pfs. The launcher enables KVM acceleration, and the benchmark driver verifies this precondition before every unikernel run and aborts otherwise.
Randomness is provided as a boot-time seed on the kernel command line, drawn from the host's \texttt{/dev/urandom}.
% (Unikraft/setup.sh, generated qemu wrappers; E4, E15.)
The guest memory allocation was raised from \SI{64}{\mebi\byte} after out-of-memory failures during porting.
% (Report/notes.txt l.266: "64M is not enough memory, increased to 128M".)

\subsection{Benchmark Orchestration}
Each measurement campaign is orchestrated for reproducibility with a Python driver. 
The driver executes all stages and metrics over all environments and algorithm configurations.
It records per-operation statistics as JSON, and stores per-algorithm timestamps for aligning the external power trace.
% (run_benchmarks.py; results JSONs in epqciuoe/Benchmark/.)
% The framework, including all build scripts and patches, is available as an
% artifact.
% \todo{Anonymized artifact link for submission; the repository currently
% resides on an institution-internal GitLab instance.}

% Evaluation (REVISED 2026-08-18 for the 2026-08-12/13/17 campaign).
% Primitive speed/memory/energy now come from three separate result files
% (see scripts/make_tables.py and scripts/{algorithms,memory,power}.yaml):
%   results/results_speed_2026-08-12_21-05-00.json  (20 s/op; standardized
%     ML-KEM/ML-DSA/SLH-DSA names alongside the legacy round-3 names;
%     classical baselines via OpenSSL EVP)
%   results/results_memory_2026-08-13_12-26-35.json (RSS, separate campaign)
%   results/results_power_2026-08-17_35-23.json      (HMC8043 PSU, separate
%     campaign; gated behind power.yaml)
% 179 (algorithm, operation) measurements across the families in
% Table~\ref{tab:family}; the appendix contains the full data.
% TLS numbers unchanged: results/tls_speed.json + results/tls_power.json
% (earlier campaign, FNB58) -- the campaign split is stated in methodology
% and limitations.
% Every number is computed directly from the raw result JSONs (see
% paper/scripts/make_tables.py). The tables in pqc-vs-unikernel/10_paper/
% remain unused (unikernel-column generator bug, E14).
\section{Evaluation}
\label{sec:evaluation}

We first evaluate the primitives in isolation (RQ1), then TLS handshakes (RQ2), and finally power and energy for both (RQ3).
Throughout, \emph{Ntv.}, \emph{Cnt.}, and \emph{Uk.}\ denote the native, container, and unikernel environments.
This section showcases the results, while Section~\ref{sec:discussion} will go into more detail on potential root causes.

\subsection{Primitive Performance}

\begin{figure*}[htb]
  \includegraphics[width=\textwidth]{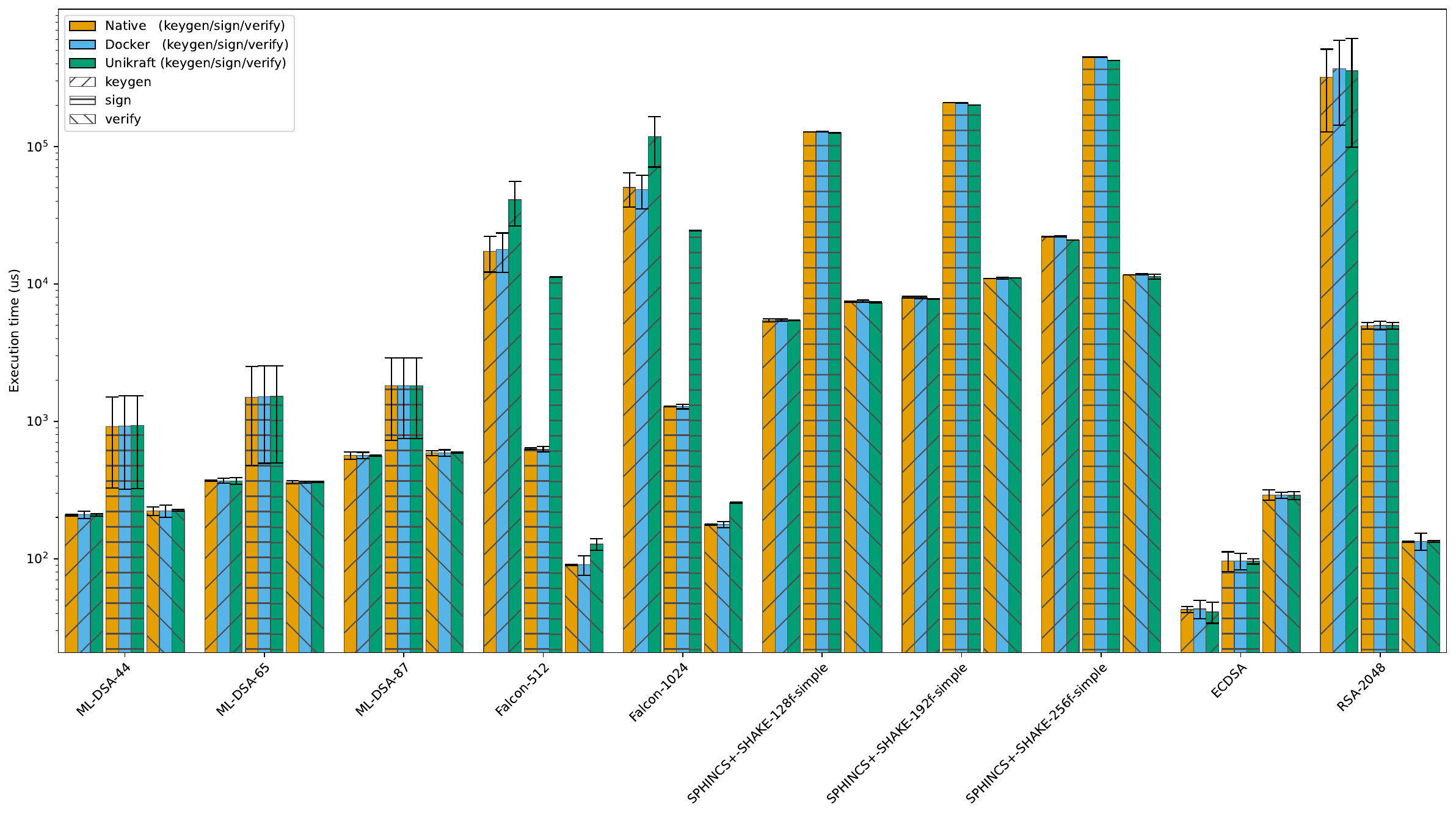}
  \caption{\label{fig:primitives_speed_sig}Mean time per operation for the signature primitives (NIST level 1 parameter sets and classical baselines) across the native, container, and unikernel environments.}
\end{figure*}

\begin{figure*}[htb]
  \includegraphics[width=0.9\textwidth]{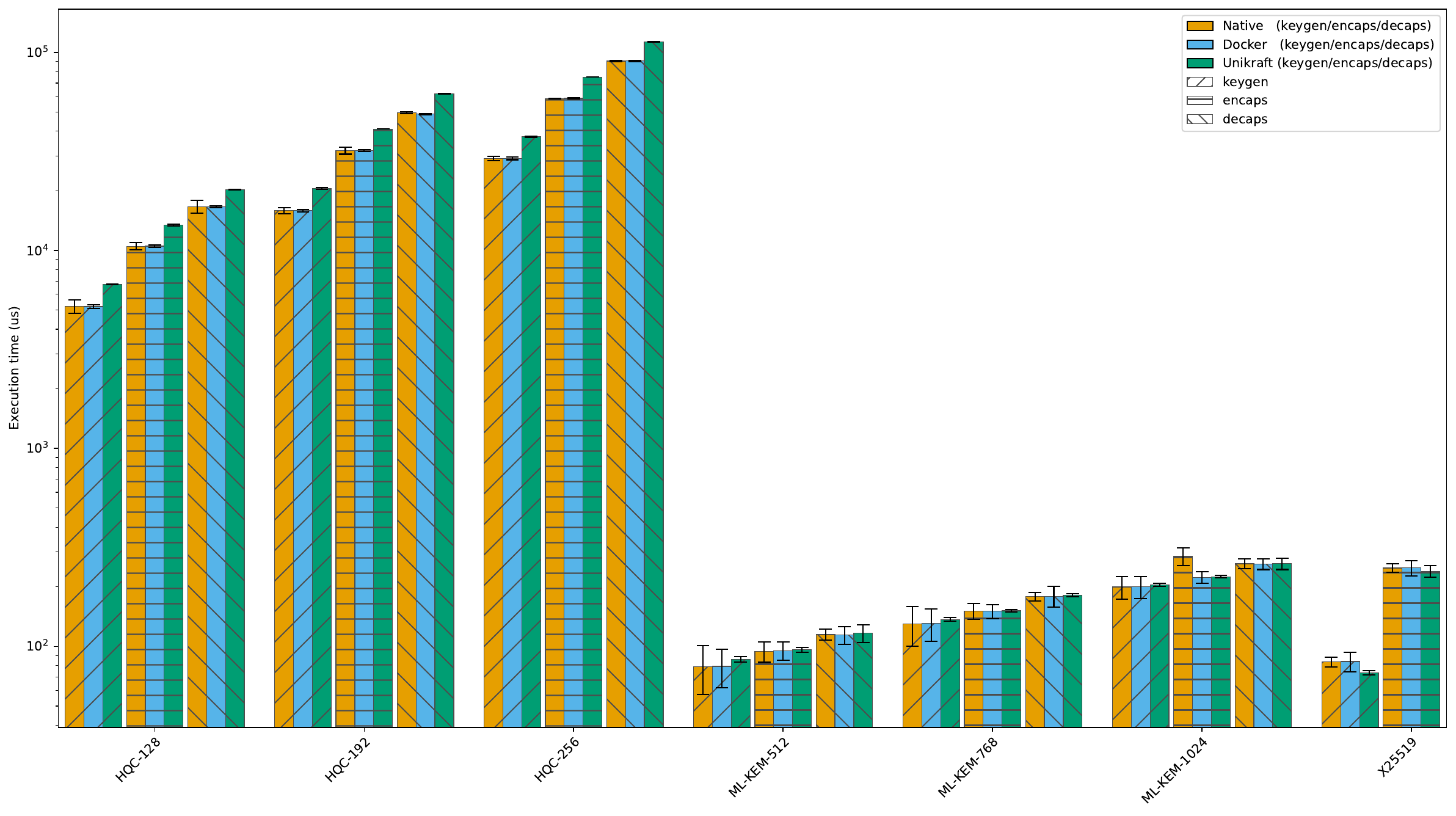}
  \caption{\label{fig:primitives_speed_kem}Mean time per operation for the KEM primitives (NIST level 1 parameter sets and classical baseline) across the native, container, and unikernel environments.}
\end{figure*}

% TODO Wording verbessern
We present a comparative assessment of the mean time per operation for the approximately 70 algorithm variants and parameter sets.
\autoref{fig:primitives_speed_sig} plots the results for the different environments for ML-DSA, Falcon, SPHINCS+ at the different security levels plus classical algorithms as a baseline.
In line with expectations, the time increases at higher security levels. Matching previous work, only ML-DSA and Falcon verification are competitive with ECDSA.
Similarly, \autoref{fig:primitives_speed_kem} shows that only ML-KEM is on par with X25519.

A representative overview of level-1 \gls{dsa} and \gls{kem} as well as a comparison between the environments is shown in Tables~\ref{tab:prim-sig} and~\ref{tab:prim-kem} (a complete set can be found in Tables~\ref{tab:full-sig} and~\ref{tab:full-kem}).
\autoref{tab:family} illustrates an aggregated overview per algorithm family to identify more quickly algorithms of interest.
When comparing container against native the overhead ranges from \num{0.96} to \num{1.03} for the \gls{dsa}, except for the SPHINCS+-SHA2 family (min \num{0.895}) with the effect being more dominant for the \textit{short} version and the higher security levels. For the baseline classical algorithms an overhead of (\num{1.155}) is restricted solely to the RSA key generation.
For \glspl{kem} the overhead ranges from \num{0.986} to \num{1.038}. The sole outliers are Classic-McEliece-460896 key generation (\num{0.870}) and ML-KEM-1024 encapsulation (\num{0.781}).
With the exception of isolated outliers, the data indicates that the container overhead remains negligible at our benchmark resolution. 
The native benchmarks reveal identical trends.
Specifically, \gls{dsa} ML-DSA offers a balanced performance profile, whereas Falcon serves as a viable alternative for verification-intensive applications.
SPHINCS+ is an alternative solely due to its conservative security assumptions, whereas CROSS is limited to providing fast key generation.
% Computed across all algs/ops from results_speed_2026-08-12_21-05-00.json
% END TODO Wording verbessern

%%% AUTO-GENERATED by paper/scripts/make_tables.py from
%%% results_speed_2026-08-12_21-05-00.json -- DO NOT EDIT BY HAND.
%%% Ntv. column: native value (mean_us); Cnt./Uk. columns: ratio to native.
%%% table* (full width): 10 columns do not fit a single acmart column.
\begin{table*}[t]
\centering
\caption{Signature primitive performance (NIST level 1 parameter sets and classical baselines). Ntv.\ is the native mean time per operation in \unit{\micro\second}, over a \SI{20}{\second} window, Cnt.$\times$ and Uk.$\times$ are container and unikernel ratios compared to native.}
\label{tab:prim-sig}
\small
\setlength{\tabcolsep}{5pt}
\begin{tabular}{l | rrr|rrr|rrr}
\toprule
 & \multicolumn{3}{c|}{Keygen} & \multicolumn{3}{c|}{Sign} & \multicolumn{3}{c}{Verify} \\
Algorithm & Ntv. & Cnt.$\times$ & Uk.$\times$ & Ntv. & Cnt.$\times$ & Uk.$\times$ & Ntv. & Cnt.$\times$ & Uk.$\times$ \\
\midrule
ML-DSA-44 & 208.4 & 1.00 & 1.00 & 914.7 & 1.01 & 1.02 & 223.2 & 1.00 & 1.01 \\
Falcon-512 & 17237.5 & 1.03 & 2.38 & 633.0 & 0.99 & 17.75 & 90.3 & 1.00 & 1.41 \\
SPHINCS+-SHA2-128f-simple & 5366.6 & 0.97 & 0.59 & 125512.5 & 0.98 & 0.59 & 7238.9 & 1.01 & 0.61 \\
SPHINCS+-SHA2-128s-simple & 345295.9 & 0.97 & 0.59 & 2629678.9 & 0.97 & 0.58 & 2640.6 & 0.92 & 0.60 \\
SPHINCS+-SHAKE-128f-simple & 5439.8 & 1.00 & 1.00 & 127668.4 & 1.00 & 0.98 & 7426.7 & 1.01 & 0.99 \\
SPHINCS+-SHAKE-128s-simple & 352488.2 & 1.00 & 0.95 & 2677162.1 & 1.00 & 0.95 & 2590.8 & 1.00 & 0.91 \\
MAYO-1 & 1070.2 & 1.00 & 1.60 & 2380.7 & 1.00 & 1.20 & 728.7 & 1.02 & 2.00 \\
cross-rsdp-128-fast & 56.8 & 1.00 & 0.98 & 1821.3 & 1.01 & 1.04 & 1037.1 & 1.00 & 1.03 \\
\hdashline
ECDSA & 42.7 & 1.01 & 0.96 & 96.5 & 1.00 & 0.99 & 291.4 & 1.00 & 0.99 \\
RSA-2048 & 319292.4 & 1.16 & 1.11 & 4967.0 & 1.00 & 1.00 & 133.1 & 1.01 & 1.01 \\
\bottomrule
\end{tabular}
\end{table*}

%%% AUTO-GENERATED by paper/scripts/make_tables.py from
%%% results_speed_2026-08-12_21-05-00.json -- DO NOT EDIT BY HAND.
%%% Ntv. column: native value (mean_us); Cnt./Uk. columns: ratio to native.
%%% table* (full width): 10 columns do not fit a single acmart column.
\begin{table*}[t]
\centering
\caption{KEM primitive performance (NIST level 1 parameter sets, classical baseline, and hybrid). Ntv.\ is the native mean time per operation in \unit{\micro\second}, over a \SI{20}{\second} window, Cnt.$\times$ and Uk.$\times$ are container and unikernel ratios compared to native.}
\label{tab:prim-kem}
\small
\setlength{\tabcolsep}{5pt}
\begin{tabular}{l | rrr|rrr|rrr}
\toprule
 & \multicolumn{3}{c|}{Keygen} & \multicolumn{3}{c|}{Encaps} & \multicolumn{3}{c}{Decaps} \\
Algorithm & Ntv. & Cnt.$\times$ & Uk.$\times$ & Ntv. & Cnt.$\times$ & Uk.$\times$ & Ntv. & Cnt.$\times$ & Uk.$\times$ \\
\midrule
ML-KEM-512 & 79.0 & 1.01 & 1.09 & 94.4 & 1.01 & 1.02 & 114.8 & 1.00 & 1.02 \\
BIKE-L1 & 42923.7 & 1.00 & 1.28 & 2176.9 & 1.00 & 1.29 & 35255.4 & 1.00 & 1.24 \\
HQC-128 & 5216.3 & 1.00 & 1.29 & 10495.6 & 1.00 & 1.28 & 16640.0 & 1.00 & 1.22 \\
FrodoKEM-640-AES & 17010.6 & 1.00 & 0.92 & 17226.8 & 1.00 & 0.92 & 17300.3 & 0.99 & 0.91 \\
FrodoKEM-640-SHAKE & 5810.3 & 1.00 & 0.99 & 6573.4 & 0.99 & 0.99 & 6483.2 & 1.00 & 1.00 \\
Classic-McEliece-348864 & 344534.1 & 1.04 & 1.06 & 168.8 & 1.00 & 1.51 & 56628.3 & 1.00 & 0.97 \\
\hdashline
ECDHE & 62.3 & 0.99 & 1.00 & 245.3 & 1.00 & 1.01 & --- & --- & --- \\
\hdashline[1pt/1pt]
x25519\_Kyber512 & 132.4 & 1.00 & 1.23 & 303.8 & 1.00 & 1.14 & 294.6 & 1.00 & 1.22 \\
\bottomrule
\end{tabular}
\end{table*}

%%% AUTO-GENERATED by paper/scripts/make_tables.py from
%%% results_speed_2026-08-12_21-05-00.json -- DO NOT EDIT BY HAND.
%%% Per-family overhead ratios across all parameter sets and operations.
\begin{table}[t]
\centering
\caption{Per-family execution-time overhead relative to native, across all parameter sets and operations of each family ($n$ = number of operation). \acrshortpl{dsa} above, \acrshortpl{kem} below.}
\label{tab:family}
\small
\setlength{\tabcolsep}{4pt}
\begin{tabular}{l r rr rr}
\toprule
 & & \multicolumn{2}{c}{Container/native} & \multicolumn{2}{c}{Unikernel/native} \\
Family & $n$ & med. & range & med. & range \\
\midrule
ML-DSA & 9 & 1.003 & 1.000--1.014 & 1.007 & 0.993--1.020 \\
Falcon & 12 & 1.000 & 0.964--1.031 & 2.307 & 1.404--19.200 \\
SPHINCS+ (SHA-2) & 18 & 0.964 & 0.895--1.012 & 0.588 & 0.542--0.671 \\
SPHINCS+ (SHAKE) & 18 & 1.003 & 0.994--1.017 & 0.959 & 0.909--1.024 \\
MAYO & 12 & 0.999 & 0.971--1.020 & 1.534 & 1.181--2.240 \\
CROSS & 45 & 1.001 & 0.981--1.014 & 1.017 & 0.965--1.088 \\
\hdashline
ECDSA & 3 & 1.001 & 0.995--1.011 & 0.993 & 0.961--0.994 \\
RSA-2048 & 3 & 1.006 & 1.003--1.155 & 1.007 & 1.003--1.114 \\
\midrule
ML-KEM & 9 & 1.001 & 0.781--1.009 & 1.016 & 0.787--1.091 \\
BIKE & 9 & 0.999 & 0.995--1.001 & 1.280 & 1.232--1.294 \\
Classic McEliece & 12 & 1.001 & 0.870--1.038 & 1.052 & 0.907--1.655 \\
HQC & 9 & 1.000 & 0.986--1.002 & 1.282 & 1.220--1.292 \\
FrodoKEM (AES) & 9 & 0.999 & 0.995--1.003 & 0.918 & 0.910--0.923 \\
FrodoKEM (SHAKE) & 9 & 0.995 & 0.994--1.007 & 0.993 & 0.984--0.998 \\
\hdashline
ECDHE & 2 & 0.998 & 0.993--1.004 & 1.002 & 0.999--1.006 \\
\bottomrule
\end{tabular}
\end{table}

For the unikernel, a more nuanced picture emerges.
ML-DSA, the SHAKE variants of SPHINCS+, and CROSS variants performed near native (\numrange{0.909}{1.088}).
For the \glspl{kem} the SHAKE variant of FrodoKEM, and largely ML-KEM perform similar to the native run (\numrange{0.986}{1.091}).
ML-KEM-1024 encapsulation (\num{0.787}) is the outlier here too, with almost exactly the same value as in the container environment.
MAYO (\num{1.534}), BIKE (\num{1.280}), and HQC (\num{1.282}) show a moderate overhead.
Falcon key generation (\num{2.304}) and verification (\num{1.427}) are in a similar range, however, signing is significantly slower with an overhead ratio from \SIrange{17.748}{19.2}{\times}.
The padded variants are also consistently slower.
Conversely, the SHA-2 variants of SPHINCS+ run consistently faster in the unikernel than natively (\numrange{0.542}{0.671}).
%
% \paragraph{Unikernel overhead is algorithm-specific.}
% The per-family view shows three regimes:
% \begin{itemize}
% \item \emph{Near-native} (median $0.92$--$1.05\times$): ECDSA, RSA, ECDHE,
%   CROSS, \emph{and the three standardized NIST post-quantum families}---
%   ML-KEM ($1.02$), ML-DSA ($1.01$), and the SHAKE variants of
%   SPHINCS+ ($0.96$)---together with Classic McEliece ($1.05$) and
%   FrodoKEM (both variants, with the AES variant slightly \emph{faster} at
%   $0.92\times$). Classic McEliece's family median is pulled up specifically
%   by encapsulation, which runs $1.5$--$1.65\times$ slower in the unikernel
%   across all four parameter sets while keygen and decapsulation stay
%   near-native.
% \item \emph{Moderate overhead} (median $1.28$--$1.53\times$): BIKE and HQC
%   ($1.28$ each) and MAYO ($1.53$).
% \item \emph{Outliers}: Falcon signing runs $17.8\times$ (Falcon-512) to
%   $19.2\times$ (Falcon-padded-1024) slower in the unikernel, while its
%   verification is only $\approx 1.4$--$1.45\times$ slower. Conversely, the
%   SHA-2 variants of SPHINCS+ run consistently \emph{faster} in the
%   unikernel than natively ($0.54$--$0.67\times$).
% \end{itemize}
%
Notably, the lattice-based algorithm families NIST has standardized, ML-KEM and ML-DSA, are among the ones that virtualize the best.
Similarly, SHAKE variants of SPHINCS+ and FrodoKEM perform even slightly faster.

\paragraph{Memory}
Memory-wise, it is important to understand that the metrics are not directly comparable, as the memory layout is different for each virtualization technology.
Instead, they provide insights into how much memory has to be allocated for each technology in practice.
Table~\ref{tab:prim-mem} shows the mean resident memory during primitive execution for the security level 1 variants of the \glspl{dsa} and \glspl{kem}.
% note the caption's caveat that the columns measure different quantities
The native processes use on average \SI{4.200}{\mebi\byte} of RAM, the containerized process image \SI{13.715}{\mebi\byte}, and the unikernel \SI{57.655}{\mebi\byte} when computing the \gls{pq} algorithms.
This shows the expected tradeoff for the statically compiled unikernel.
While the classical algorithms require similar amounts of memory in the container, and even slightly less in the native environment, the unikernel variant requires almost \SI{25}{\percent} more, when running the OpenSSL-implemented classical baselines.
This is again expected based on the increased dependencies.
Within one environment, differences between \gls{pqc} algorithms are small.
The maximum difference between signature algorithms is ~\SI{0.999}{\mebi\byte} and \SI{1.591}{\mebi\byte} for \glspl{kem}.
Both this difference and the standard deviation are negligible between all environments, with only containers showing a slightly less spread ($\sigma_{cnt}=0.145$ compared to $\sigma_{nat}=0.296$ and $\sigma_{uk}=0.301$).
In the native environment the SHAKE variant of SPHINCS+ requires the least memory.
However, it has some of the outliers with the highest memory demand in the container environment.
For unikernel it requires only slightly less memory than the mean.
MAYO tends to require the most memory between all environments.
For \glspl{kem} Classic-McEliece requires more than average memory.
BIKE, ML-KEM, and FrodoKEM tend to require less than average memory across all environments and operations.
However, all have a few outliers, often in decapsulation in the container environment, where these algorithms require more memory.

%%% AUTO-GENERATED by paper/scripts/make_tables.py from
%%% results_memory_2026-08-13_12-26-35.json (memory_rss_avg_kb, averaged over the
%%% category's operations) -- DO NOT EDIT BY HAND.
\begin{table}[t]
\centering
\caption{Mean resident memory (\unit{\mebi\byte}) during primitive execution (\acrshortpl{dsa} above, \acrshortpl{kem} below), averaged over all operations of each algorithm.}
\label{tab:prim-mem}
\small
\begin{tabular}{l | rrr}
\toprule
Algorithm & Native & Container & Unikernel \\
\midrule
ML-DSA-44 & 4.1 & 13.7 & 57.4 \\
Falcon-512 & 4.1 & 13.6 & 57.5 \\
SPHINCS+-SHA2-128f-simple & 4.1 & 13.6 & 57.5 \\
SPHINCS+-SHA2-128s-simple & 4.0 & 13.7 & 57.6 \\
SPHINCS+-SHAKE-128f-simple & 3.9 & 13.7 & 57.5 \\
SPHINCS+-SHAKE-128s-simple & 3.9 & 13.7 & 57.6 \\
MAYO-1 & 4.5 & 13.7 & 57.8 \\
cross-rsdp-128-fast & 4.0 & 13.6 & 57.5 \\
\hdashline
ECDSA & 4.0 & 13.7 & 76.1 \\
RSA-2048 & 3.9 & 13.6 & 78.5 \\
\midrule
ML-KEM-512 & 4.0 & 13.7 & 57.4 \\
BIKE-L1 & 3.9 & 13.7 & 57.4 \\
HQC-128 & 4.1 & 13.7 & 57.5 \\
FrodoKEM-640-AES & 3.9 & 13.9 & 57.5 \\
FrodoKEM-640-SHAKE & 3.9 & 13.7 & 57.5 \\
Classic-McEliece-348864 & 4.5 & 13.6 & 58.0 \\
\hdashline
ECDHE & 4.0 & 13.7 & 76.4 \\
\bottomrule
\end{tabular}
\end{table}

\subsection{TLS Handshake Throughput}

Table~\ref{tab:tls-speed} reports our results for completed initial TLS~1.3 handshakes in a 30\,s window.
As discussed in Section~\ref{sec:methodology}, the client environments reach the always native server over different network paths inside the \gls{pi4}, so the numbers reflect deployment-level differences including transport, not cryptographic overhead alone.
They also stem from an earlier measurement campaign than the primitive results.

%%% AUTO-GENERATED by paper/scripts/make_tables.py from
%%% epqciuoe/Benchmark/results/tls_speed.json -- DO NOT EDIT BY HAND.
%%% Average wall-clock time per completed initial TLS 1.3 connection, derived as
%%% (30 s window) / (connections completed within the window), client side,
%%% server always native (see methodology). Ratios to native in parentheses.
%%% NOTE: TLS data stems from the earlier (pre-2026-08 primitives) campaign.
%%% SPHINCS+-128s abbreviates SPHINCS+-SHA2-128s-simple; Frodo-640 abbreviates FrodoKEM-640.
\begin{table}[t]
\centering
\caption{Mean time per completed initial TLS~1.3 handshake in \unit{\milli\second}, divided by the number of handshakes completed in the \SI{30}{\second}  measurement window. Server native environment in all cases. Ratios to native in parentheses.}
\label{tab:tls-speed}
\footnotesize
\setlength{\tabcolsep}{3pt}
\begin{tabular}{ll | rrr}
\toprule
Signature & KEM & Native & Container & Unikernel \\
\midrule
Dilithium2 & Kyber512 & 5.0 & 6.3 (1.25) & 6.6 (1.31) \\
Falcon-512 & Kyber512 & 5.1 & 6.4 (1.26) & 7.0 (1.36) \\
Dilithium3 & Kyber768 & 5.7 & 7.2 (1.27) & 7.2 (1.28) \\
Falcon-1024 & Kyber768 & 5.9 & 7.4 (1.26) & 7.7 (1.30) \\
Dilithium3 & Kyber1024 & 5.8 & 7.3 (1.25) & 7.4 (1.28) \\
Falcon-1024 & Kyber1024 & 6.1 & 7.6 (1.25) & 7.9 (1.30) \\
SPHINCS+-128s & Kyber512 & 2727 & 2727 (1.00) & 2727 (1.00) \\
Dilithium2 & BIKE-L1 & 156 & 143 (0.91) & 158 (1.01) \\
Dilithium2 & HQC-128 & 50.2 & 48.2 (0.96) & 51.7 (1.03) \\
Dilithium2 & Frodo-640-AES & 71.1 & 74.3 (1.04) & 70.4 (0.99) \\
Dilithium2 & Frodo-640-SHAKE & 27.9 & 30.3 (1.08) & 30.0 (1.07) \\
\hdashline
RSA-2048 & ECDHE & 9.5 & 11.0 (1.16) & 11.3 (1.19) \\
ECDSA & ECDHE & 5.1 & 6.5 (1.29) & 6.9 (1.36) \\
\bottomrule
\end{tabular}
\end{table}

% \paragraph{Fast algorithm combinations expose the environment.}
The lattice-based combinations (Dilithium/Falcon + Kyber) perform the best among all tested combinations and throughout all environments.
For the container the slowdown for these combinations is \SIrange{25}{27}{\percent} and for the unikernel environment \SIrange{28}{36}{\percent}.
For our setup Dilithium2+Kyber512 performed the best among all combinations, with \num{5.0} seconds per connection in the native environment and \num{6.3} and \num{6.6} in container and unikernel.
That means that this pairing is on par with classical ECDSA+ECDHE in every environment, which matches prior observations that lattice-based \gls{pqc} is competitive with classical elliptic-curve cryptography on handshake cost~\cite{Tasopoulos2022}.
When looking at the slower combinations, it becomes obvious that expensive cryptography hides the environment.
This is especially obvious for SPHINCS+ where the expensive verification operation limits all environments to \num{11} connections in all three environments (\num{2.7} seconds per connection).
At this scale the per-connection transport and virtualization cost is negligible relative to the cryptographic cost.
Interestingly, for BIKE-L1 and HQC-128 the container environment outperforms the native one.
However, the difference is within plausible single-run variation, which is a plausible explanation as we did not see such behavior in our primitive runs.

\subsection{Power and Energy}
We conducted a comprehensive evaluation of power and energy, as it is often a limiting factor for embedded devices.

\paragraph{Primitives}
Table~\ref{tab:prim-energy} reports the power and energy per operation for each algorithm family, averaged across all of the family's parameter sets, integrated from the HMC8043 power trace, with the device's idle baseline power subtracted out (see Section~\ref{sec:methodology}).
Raw values are listed in Table~\ref{tab:full-energy-sig} and \ref{tab:full-energy-kem}.
Mean power during \gls{pqc} primitive execution ranges from \SIrange{0.97}{1.57}{\watt} across algorithms and environments.
However, execution time is still the driving factor so that the overall baseline-subtracted energy results still closely track the time results.
ML-DSA shows good performance overall.
However, it does not achieve the very low energy requirements of ECDSA.
Falcon requires the least energy ($E_{nat}=\SI{0.166}{\milli\joule}$) for verification and CROSS ($E_{nat}=\SI{0.100}{\milli\joule}$) the least for key generation.
SPHINCS+, on the other hand, requires three to four orders of magnitude more than the lattice schemes.
E.g., SPHINCS+ SHA-2 signing requires \SI{1594}{\times} more energy than ML-DSA in the native environment.
Similarly, for the \glspl{kem} ML-KEM has an average power draw of \SI{1.21}{\watt} in the native environment.
Its fast execution drops the required energy to \SI{1.93}{\milli\joule}, which is on equal footing with ECDHE.
In contrast, HQC requires more power in all but the unikernel environment than ML-KEM, which confounds the effect of the longer execution time per operation.
This means it requires two orders of magnitude more energy in the end.

The environments seem to have only a marginal impact on the required power overall.
The average container to native ratio over all \gls{pq} \glspl{dsa} is \SI{1.021}{\times} and the unikernel to native ratio is \SI{0.980}{\times}.
For the \glspl{kem} the values are \SI{1.023}{\times} and \SI{0.958}{\times}.
While still small in absolute numbers, the unikernel thus reduces the required power.
When comparing energy between environments, execution time again becomes the major contributor.
Even though Falcon's power draw drops from \SI{1.1}{\watt} to \SI{0.97}{\watt} when shifting from native to unikernel execution, its energy rises by a factor of \SI{16.8}{\times} to \SI{17.82}{\milli\joule}.
This is consistent with the overhead during speed measurements of \SI{17.75}{\times} for Falcon-512 signing.
Similarly, the SHA-2 variant of SPHINCS+ is able to reduce its average above baseline power cost from \SI{1.28}{\watt} in the native environment to \SI{1.24}{\watt} in the unikernel (factor: \SI{0.974}{\times}).
Paired with the previously observed speed-up this results in a reduction of the required energy by a factor of \SI{0.60}{\times}.
% 1,083.51666
% 1,809.04
In conclusion the measurements show that energy per operation tracks execution time.
% (e_per_iteration/p_per_iteration/baseline_mw fields; consistency e ≈ P·t
%  verified in the branch-delta audit; e_marginal = e·(p-baseline)/p, see
%  inject_marginal_energy() in scripts/make_tables.py.)

%%% AUTO-GENERATED by paper/scripts/make_tables.py from
%%% results_power_2026-08-17_35-23.json -- DO NOT EDIT BY HAND.
%%% Merged signature+KEM energy table, one row per (family, operation) so it
%%% fits a single acmart column. Each row is the mean over all of the family's
%%% parameter sets (FAMILIES/FAMILY_ALGS), not a single NIST level~1 representative.
%%% $E$ ('e_marginal_per_iteration') and $P$ (p_marginal_per_iteration) are BOTH baseline-subtracted: the
%%% operation's energy/power cost above the ambient idle baseline, not the raw
%%% e_per_iteration/p_per_iteration (see inject_marginal_energy() -- baseline drift is
%%% otherwise large enough relative to many operations' own power draw to flip
%%% container/unikernel-vs-native ratios). $P$ is in W, $E$ in mJ.
%%% All three environment columns give the plain family-mean absolute value (no
%%% ratio-to-native -- dropped per user request to fit the table's width; see
%%% table-family.tex for container/unikernel-vs-native ratios, and table-power-full.xlsx
%%% for P-Baseline with ratios).
\begin{table}[t]
\centering
\caption{Mean power $P$ (\unit{\watt}) and energy $E$ (\unit{\milli\joule}) per operation above the idle baseline (\acrshort{dsa} and \acrshort{kem} primitive families and classical baselines), averaged across all of each family's NIST parameter sets. Derived from the HMC8043 power trace aligned with per-operation timestamps.}
\label{tab:prim-energy}
\footnotesize
\setlength{\tabcolsep}{2.5pt}
\begin{tabular}{ll rr|rr|rr}
\toprule
 & & \multicolumn{2}{c|}{Native} & \multicolumn{2}{c|}{Container} & \multicolumn{2}{c}{Unikernel} \\
Family & Op & $P$ & $E$ & $P$ & $E$ & $P$ & $E$ \\
 & & (W) & (mJ) & (W) & (mJ) & (W) & (mJ) \\
\midrule
\multirow{3}{*}{ML-DSA} & Keygen & 1.23 & 0.485 & 1.17 & 0.467 & 1.20 & 0.446 \\
 & Sign & 1.22 & 1.83 & 1.17 & 1.72 & 1.31 & 1.92 \\
 & Verify & 1.18 & 0.475 & 1.25 & 0.519 & 1.30 & 0.518 \\
\multirow{3}{*}{Falcon} & Keygen & 1.07 & 33.78 & 1.15 & 40.29 & 1.02 & 84.50 \\
 & Sign & 1.10 & 1.06 & 1.21 & 1.21 & 0.97 & 17.82 \\
 & Verify & 1.22 & 0.166 & 1.17 & 0.165 & 1.07 & 0.211 \\
\multirow{3}{*}{SPHINCS+ (SHA-2)} & Keygen & 1.27 & 323 & 1.19 & 288 & 1.20 & 183 \\
 & Sign & 1.33 & 2917 & 1.31 & 2726 & 1.29 & 1620 \\
 & Verify & 1.23 & 10.55 & 1.26 & 10.33 & 1.24 & 6.04 \\
\multirow{3}{*}{SPHINCS+ (SHAKE)} & Keygen & 1.25 & 311 & 1.20 & 287 & 1.23 & 285 \\
 & Sign & 1.21 & 2330 & 1.33 & 2784 & 1.31 & 2652 \\
 & Verify & 1.18 & 9.93 & 1.34 & 11.29 & 1.22 & 9.49 \\
\multirow{3}{*}{MAYO} & Keygen & 1.28 & 5.58 & 1.29 & 6.21 & 1.12 & 7.87 \\
 & Sign & 1.22 & 11.10 & 1.22 & 12.74 & 1.13 & 13.21 \\
 & Verify & 1.57 & 4.37 & 1.44 & 4.01 & 1.15 & 6.48 \\
\multirow{3}{*}{CROSS} & Keygen & 1.21 & 0.100 & 1.22 & 0.110 & 1.23 & 0.105 \\
 & Sign & 1.32 & 8.79 & 1.32 & 9.29 & 1.25 & 8.58 \\
 & Verify & 1.30 & 4.97 & 1.28 & 5.16 & 1.27 & 4.88 \\
\hdashline
\multirow{3}{*}{ECDSA} & Keygen & 1.10 & 0.047 & 1.29 & 0.063 & 1.20 & 0.050 \\
 & Sign & 0.93 & 0.089 & 1.12 & 0.114 & 1.09 & 0.107 \\
 & Verify & 0.89 & 0.260 & 1.06 & 0.323 & 0.95 & 0.281 \\
\multirow{3}{*}{RSA-2048} & Keygen & 0.99 & 363 & 0.73 & 261 & 1.01 & 349 \\
 & Sign & 0.98 & 4.98 & 0.68 & 3.65 & 1.01 & 5.20 \\
 & Verify & 0.75 & 0.102 & 0.69 & 0.101 & 0.97 & 0.133 \\
\midrule
\multirow{3}{*}{ML-KEM} & Keygen & 1.14 & 0.157 & 1.25 & 0.190 & 1.21 & 0.178 \\
 & Encaps & 1.23 & 0.188 & 1.30 & 0.212 & 1.25 & 0.203 \\
 & Decaps & 1.27 & 0.233 & 1.35 & 0.271 & 1.18 & 0.229 \\
\multirow{3}{*}{BIKE} & Keygen & 1.44 & 247 & 1.47 & 255 & 1.25 & 250 \\
 & Encaps & 1.40 & 12.12 & 1.23 & 11.67 & 1.11 & 13.63 \\
 & Decaps & 1.42 & 205 & 1.30 & 207 & 1.04 & 175 \\
\multirow{3}{*}{Classic McEliece} & Keygen & 1.40 & 914 & 1.49 & 921 & 1.48 & 743 \\
 & Encaps & 1.24 & 0.346 & 1.41 & 0.399 & 1.25 & 0.534 \\
 & Decaps & 1.07 & 78.48 & 1.07 & 81.12 & 1.07 & 79.48 \\
\multirow{3}{*}{HQC} & Keygen & 1.38 & 22.56 & 1.43 & 25.01 & 1.14 & 26.34 \\
 & Encaps & 1.33 & 42.26 & 1.46 & 50.27 & 1.10 & 49.98 \\
 & Decaps & 1.42 & 74.12 & 1.36 & 76.50 & 1.04 & 76.11 \\
\multirow{3}{*}{FrodoKEM (AES)} & Keygen & 1.12 & 46.86 & 1.06 & 51.46 & 1.12 & 45.91 \\
 & Encaps & 1.14 & 52.93 & 1.22 & 57.38 & 1.25 & 51.31 \\
 & Decaps & 1.11 & 51.54 & 1.14 & 52.67 & 1.25 & 52.88 \\
\multirow{3}{*}{FrodoKEM (SHAKE)} & Keygen & 1.25 & 17.38 & 1.23 & 19.41 & 1.30 & 18.53 \\
 & Encaps & 1.25 & 19.23 & 1.27 & 22.29 & 1.25 & 19.80 \\
 & Decaps & 1.39 & 21.91 & 1.25 & 21.45 & 1.23 & 19.77 \\
\hdashline
\multirow{3}{*}{ECDHE} & Keygen & 1.19 & 0.075 & 1.02 & 0.066 & 1.21 & 0.076 \\
 & Encaps & 1.23 & 0.300 & 0.92 & 0.230 & 1.05 & 0.266 \\
 & Decaps & --- & --- & --- & --- & --- & --- \\
\bottomrule
\end{tabular}
\end{table}

\paragraph{TLS: energy per handshake follows throughput.}
Table~\ref{tab:tls-power} shows mean system power and derived energy per completed handshake for the TLS campaign.
% (FNB58 meter, $\approx$62\,s windows, connection counts from the same run). 
Mean system power varies little for the \gls{pqc} algorithms (\SIrange{3.21}{3.46}{\watt}).
Energy per handshake, however, follows throughput: for the fast lattice combinations the system energy cost per handshake is \SIrange{34}{57}{\milli\joule}, outperforming RSA and in some combinations even ECDSA with ECDHE.
However, switching the \gls{kem} from the fast Kyber512 to the second fastest Frodo-640-SHAKE already increases the required system energy by a factor of \SI{5.67}{\times}.
For the \glspl{dsa}, Falcon is only slightly more energy intensive, on the other hand, SPHINCS+ requires \SI{567}{\times} more system energy.
When comparing environments, container costs more system energy per handshake than native by a factor of \SI{1.022}{\times}.
Unikernels require even more energy, raising the factor to \SI{1.039}{\times}.
However, summarizing, algorithm choice dwarfs environment choice: a SPHINCS+ signed handshake costs three orders of magnitude more energy than a Dilithium signed one in every environment.

%%% AUTO-GENERATED by paper/scripts/make_tables.py from
%%% epqciuoe/Benchmark/results/tls_power.json -- DO NOT EDIT BY HAND.
%%% Window duration derived from recorded timestamps: 62--66 s per combination.
%%% NOTE: TLS power stems from the earlier campaign (FNIRSI FNB58 meter).
\begin{table}[t]
\centering
\caption{Mean system power draw $P$ (\unit{\watt}) and energy per completed handshake $E_{hs}$ (\unit{\milli\joule}) during the TLS power campaign. Measured inline at the device supply with the FNIRSI FNB58. Connection counts and energy stem from the same run.}
\label{tab:tls-power}
\footnotesize
\setlength{\tabcolsep}{2.5pt}
\begin{tabular}{ll rr|rr|rr}
\toprule
 & & \multicolumn{2}{c|}{Native} & \multicolumn{2}{c|}{Container} & \multicolumn{2}{c}{Unikernel} \\
Signature & KEM & $P$ & $E_{hs}$ & $P$ & $E_{hs}$ & $P$ & $E_{hs}$ \\
\midrule
Dilithium2 & Kyber512 & 3.26 & 34 & 3.27 & 44 & 3.42 & 48 \\
Falcon-512 & Kyber512 & 3.24 & 35 & 3.25 & 45 & 3.37 & 50 \\
Dilithium3 & Kyber768 & 3.29 & 39 & 3.29 & 50 & 3.44 & 53 \\
Falcon-1024 & Kyber768 & 3.25 & 41 & 3.29 & 52 & 3.41 & 56 \\
Dilithium3 & Kyber1024 & 3.28 & 40 & 3.29 & 51 & 3.46 & 55 \\
Falcon-1024 & Kyber1024 & 3.28 & 42 & 3.28 & 53 & 3.43 & 57 \\
SPHINCS+-128s & Kyber512 & 3.32 & 19281 & 3.33 & 19710 & 3.41 & 19933 \\
Dilithium2 & BIKE-L1 & 3.25 & 1053 & 3.36 & 997 & 3.25 & 1073 \\
Dilithium2 & HQC-128 & 3.22 & 334 & 3.34 & 334 & 3.25 & 350 \\
Dilithium2 & Frodo-640-AES & 3.21 & 472 & 3.22 & 499 & 3.26 & 476 \\
Dilithium2 & Frodo-640-SHAKE & 3.34 & 193 & 3.35 & 217 & 3.39 & 212 \\
\hdashline
RSA-2048 & ECDHE & 3.17 & 63 & 3.21 & 75 & 3.29 & 79 \\
ECDSA & ECDHE & 3.23 & 35 & 3.24 & 45 & 3.32 & 49 \\
\bottomrule
\end{tabular}
\end{table}

% Discussion. Interprets the evaluation. The entropy hypothesis stems from the
% repository authors' own analysis (10_paper/sections/40_evaluation.tex l.63;
% Report/document.tex) and the verified seeding mechanism (Unikraft/setup.sh,
% E15); it is presented as a hypothesis, not a finding. Security trade-off
% remarks are qualitative and cited; no security experiments exist in the repo.
% The Dilithium/ML-DSA and Kyber/ML-KEM paragraph is grounded in
% epqciuoe/docs/mlkem-kyber-hybrid-findings.md (root-caused for the KEM side via
% liboqs release notes) and epqciuoe/Unikraft/libs/lib-oqs/include/oqs/oqsconfig.h
% (confirms the same *_aarch64 macro pattern for the SIG side); ratios recomputed
% from results_speed_2026-08-12_21-05-00.json, matching evidence-map.md E27/E31/E35.
% The runtime-selection cause (why the compiled-in aarch64 backend isn't used
% inside the Unikraft guest) is NOT confirmed -- flagged as \todo, since
% plan_implementation_logging.md's instrumentation was never executed.
\section{Discussion}
\label{sec:discussion}

% \paragraph{Virtualization cost is inversely relevant to cryptographic cost.}
When looking at the unikernel, we consistently observe at the primitives level that the standardized lattice-based algorithms run at essentially native speed in the unikernel.
However, MAYO, BIKE, HQC, and above all Falcon's signing routine, which we will discuss in more detail in the following paragraph, carry a measurable unikernel penalty.
At the TLS level containers and unikernels each still give up between one seventh and one quarter of native throughput for fast handshake combinations, including the standardized lattice pairing Dilithium2+Kyber512, while for BIKE, HQC, and FrodoKEM the environments are harder to tell apart and for SPHINCS+ indistinguishable.
This can be explained by the larger portion of the static costs associated with virtualization, like the different network paths in the environments (see Section~\ref{sec:limitations}).
So at least part of this TLS-level gap plausibly reflects the unikernel's connection path rather than the cryptography running over it.
The energy results reinforce this observation from a different angle.
Mean power draw differs only marginally across environments, both for primitives (\SIrange{0.96}{1.02}{\times} native, averaged over all \gls{pq} families) and for TLS handshakes (system energy per handshake at \SI{1.022}{\times} native in the container and \SI{1.039}{\times} in the unikernel).
In the cases where environment influences energy, it does so mainly by moving time rather than power.
This shows that virtualization cost is inversely related to cryptographic cost and algorithm choice remains the dominant factor.
Only when the algorithms are sufficiently fast, the choice of environment begins to influence the performance.
% Practitioners
% deploying the standardized lattice or hash-based algorithms therefore pay a
% real -- though not fully disentangled -- transport-or-cryptographic cost in
% a unikernel deployment; Falcon signing and the non-standardized alternates
% add a genuine cryptographic cost on top of that, and deployments of
% conservative or code-based schemes can otherwise choose the environment
% freely on other grounds.

% \paragraph{The Falcon anomaly.}
As discussed, Falcon signing is the single largest outlier (\SIrange{17.75}{19.20}{\times}).
Since Falcon is the only benchmarked scheme whose signing procedure depends on double-precision floating-point arithmetic~\cite{Alagic2022}, a plausible cause for this behavior includes a different floating-point/math-library configuration in the Unikraft build or emulated floating-point paths in the guest.
This is reinforced by the fact that verification, which is integer-dominated, shows only moderate overhead.
To exclude emulated execution as the cause we enforced that KVM acceleration was active during the recorded campaigns.
Notably, the anomaly is confined to latency, not power draw.
Falcon signing's mean power is in fact slightly lower in the unikernel than natively (\SI{0.97}{\watt} vs. \SI{1.10}{\watt}), a behavior more consistent with the guest stalling, e.g.\ on an emulated or absent hardware FPU path, than with it doing additional work.
However, we were not able to trace the root cause to e.g. a missing floating-point configuration or math library.
Until root-caused, Falcon signing inside Unikraft on QEMU should be considered impractical on this class of hardware.

The SPHINCS+ split is equally notable: the speed-up is specific to the SHA-2 parameter sets (\SIrange{0.54}{0.67}{\times}), while the SHAKE parameter sets perform near-native.
Since the two variants differ exactly in their internal hash function, this points at the SHA-2 code path rather than at the signature scheme itself.
The power trace corroborates that this is a genuine effect rather than measurement noise.
Mean power above baseline for the SHA-2 variant is also slightly lower in the unikernel than natively (\SI{1.24}{\watt} vs. \SI{1.28}{\watt}, factor \SI{0.974}{\times}), which combined with the speed-up compounds to a \SI{0.60}{\times} reduction in energy per operation.
This is caused by an indirection during compilation.
The unikernel uses the liboqs SHA-2 implementation instead of the one by OpenSSL, as \texttt{OPENSSL\_cpuid\_setup} normally crashes the unikernel via its SIGILL probe, and this choice also reduces build complexity.
However, this also pulls in an ARMv8 Cryptography-Extension-accelerated SHA-2, compiled with \texttt{-mcpu = cortex-a53+crypto}.
This setup is most likely faster than the OpenSSL version and thus the main contributor to the speed-up.

% \paragraph{Entropy handling and symmetric-primitive code paths.}
The unikernel receives its randomness as a one-time boot seed and expands it internally, while native and containerized processes call into the host kernel for entropy during operation.
This may also contribute to the divergent behavior to a lesser extent, as algorithms that request entropy frequently pay a syscall-path cost natively and in containers that
the unikernel avoids, whereas the unikernel's software expansion adds cost elsewhere.
This effect could be isolated by an experiment fixing the entropy source across environments.
Security-wise, boot-time-only seeding is a double-edged sword: it removes a runtime dependency, but the guest's entire entropy pool derives from one seed exposed on the QEMU command line.
This has to be considered during deployment of cryptographic unikernel workloads, as this flag is visible to host-side observers.
% The command-line visibility follows directly from the generated QEMU wrapper
% (SEED embedded in -append); the security note is an observation about the
% mechanism, not a claimed vulnerability or measured attack.

% \paragraph{Legacy vs.\ standardized naming: an implementation-selection artifact.} 
The liboqs 0.12.0 release ships both the round-3 submissions Dilithium/Kyber and the FIPS-final ML-DSA/ML-KEM used throughout the previous sections.
Still, comparing these versions shows some interesting results.
On this \gls{pi4} target, \texttt{oqsconfig.h} compiles in an AArch64-optimized backend for every Dilithium and Kyber parameter set, but no such backend exists for ML-DSA or
ML-KEM in this liboqs version, which therefore always run the portable reference implementation.
This leads to a measurable effect.
The ratios between native and container for Dilithium, ML-DSA, Kyber, and ML-KEM and the ratio for ML-DSA and ML-KEM in the unikernel are all approximately \SI{1.0}{\times}.
However, it rises, on average, to \SI{1.568}{\times} for Dilithium and \SI{2.220}{\times} for Kyber in the unikernel environment.
This is not caused by a slowdown of the FIPS versions, but instead of a faster runtime of the round-3 variants in both the native and container environment.
For example, Dilithium2 signing takes \SI{466.618}{\micro\second}, but ML-DSA-44 requires \SI{914.715}{\micro\second} in the native environment. 
Dilithium2 needs only around \SI{0.5}{\times} the time.
While the container has the same \SI{50}{\percent} difference, in the unikernel Dilithium needs \SI{937.512}{\micro\second} and ML-DSA \SI{933.227}{\micro\second}, basically the same amount of time.
This shows an environment-specific slowdown inside the Unikraft guest: even though the optimized \texttt{\_aarch64} code is compiled into that same guest image, it is evidently not being selected at runtime there.
Plausible mechanisms for this behavior include liboqs's runtime CPU-feature probe reporting negatively, or being unavailable, inside Unikraft's minimal runtime despite the feature being physically present, or a mismatch between the vCPU model QEMU presents to the guest and the host CPU used for native/container.

% Hybrid Algorithms
We evaluated with X25519-Kyber also a hybrid \gls{kem}.
However, as expected, its performance closely follows its individual component algorithms with an overhead of $\approx\SI{1.2}{\percent}$ in the native environment.
The environment speed overhead lies between the individual overheads with no apparent additional influence.
Memory is additive for native, but only up to $\approx\SI{56}{\percent}$ of the individual parts. % Ntv: 4.4 / (3.8+4.0) = 0.564
Due to the larger size, this effect is marginal for container and unikernel.
Energy is dominated by the execution time, leading to the same results as for the speed performance overhead.
Due to its composition, it follows Kyber and not ML-KEM, which leads to slightly different splits.     % kinda obvious aber mit den figures bin ich selber mal drauf reingefallen
%

% \todo{Root cause of the guest-only fallback is untested;
% plausible mechanisms include liboqs's runtime CPU-feature probe (e.g., an
% \texttt{AT\_HWCAP}/NEON check) reporting negatively, or being unavailable,
% inside Unikraft's minimal runtime despite the feature being physically
% present, or a mismatch between the vCPU model QEMU presents to the guest
% and the host CPU used for native/container. Confirming this needs
% per-operation implementation-selection logging (planned but never executed;
% see \texttt{plan\_implementation\_logging.md}), and is out of scope here.
% Practically, this means Kyber/Dilithium and ML-KEM/ML-DSA numbers should
% never be compared against each other as if they measured the same
% algorithm under different build settings; upgrading to liboqs $\geq 0.13.0$,
% which adds a formally verified AArch64 ML-KEM backend, would likely close
% most of this particular gap, but was not evaluated in this campaign.}

% \paragraph{Container vs.\ unikernel.}
When comparing container with unikernel, for pure computation the container is essentially free while the unikernel is not, the opposite of what one might expect given that a unikernel eliminates system-call and scheduling overhead.
For power, the situation reverses and the unikernel's mean power draw during primitive execution is, if anything, slightly below native, so its computational cost shows up in execution time and only marginally in power.
For TLS, the two are close, but especially for the fast algorithms the container consistently beats the unikernel environment.
The distinct path to the server for each environment may confound this observation.
Memory tells another story at a different layer: the unikernel image adds about \SI{5}{\times} the amount of resident memory over native compared to the container environment as seen from the host.
The unikernel figure includes the QEMU process itself.
% The unikernel figure includes the QEMU process itself and is measured against a fixed \SI{1}{\gibi\byte} guest allocation for this benchmark leg (\SI{128}{\mebi\byte} for TLS wrapper's allocation, not this one). -- dropped: guest allocation figure not directly relevant here and conflicts with the incomplete "raised from 64 MiB" sentence in 05-experimental-setup.tex.
The way we measure the memory of the container makes it, however, largely independent of the selected algorithm (see Section~\ref{sec:limitations}).
% Which of the two, environment or algorithm, drives this footprint also differs by layer: per algorithm, native memory usage is essentially uncorrelated with the container's ($n=70$, Pearson $r\approx0.22$). 
% Read causally this would suggest the container's fixed overhead swamping any algorithm-specific difference, but the container figure is not actually the containerized benchmark process's own memory -- the driver tracks the host-side \texttt{containerd-shim} process instead, whose footprint is largely independent of the algorithm running inside the container regardless of any real virtualization overhead. The near-zero correlation is therefore at least as likely a measurement-target artifact as a statement about container overhead.
On the other hand native memory correlates moderately with the unikernel's (Spearman $\rho\approx0.44$--$0.49$, $p<0.01$ for signatures and KEMs alike), so some of each algorithm's native memory footprint survives underneath the unikernel's much larger fixed cost.
However, as the other discussions in this section show, these effects are only marginal compared to when differences arise in the choice of implementation, either during build or runtime.
% All numbers from Tables in section 6; comparison framing only, no external
% claims. Isolation-strength comparison intentionally qualitative:
Finally, our study measures performance only and does not discuss the direct security considerations.
The isolation guarantees of containers and unikernels differ qualitatively~\cite{Madhavapeddy2014,Kuenzer2021}, and a deployment decision must weigh both.

% Limitations. Every item corresponds to a verified gap in the repository
% (evidence-map.md E8, E17, E18, E23, E24; assumptions-and-todos.md T3-T5,
% T9-T12). Stating these openly is required by the no-hallucination policy and
% strengthens reviewer trust.
\section{Limitations}
\label{sec:limitations}

% \paragraph{Virtualization mode now enforced and logged.}
The underlying virtualization method has a large impact on the performance of a unikernel. 
The unikernel launcher derives its \texttt{-enable-kvm} flag from whether \texttt{/dev/kvm} exists and falls back silently. 
To control this behavior the benchmark driver precedes every unikernel run with an explicit check that aborts the whole campaign if \texttt{/dev/kvm} is missing or not read/write accessible. 
Thus, all three primitive-campaign runs underlying this paper's numbers confirm KVM was active, allowing us to rule out TCG-emulation as a possible cause for outliers.
% This control post-dates the earlier TLS
% campaign (Section~\ref{sec:methodology}), for which no equivalent driver
% log exists and KVM status remains unconfirmed.

% \paragraph{Asymmetric network paths in the TLS benchmark.}
The TLS server always runs natively on the \gls{pi4}.
The clients connect over different stacks, depending on the environment: loopback (native), Docker's host gateway (container), or a QEMU bridge (unikernel). 
The measured differences on fast combinations therefore conflate handshake computation with network-path cost, thus the absolute throughput ratios should not be read as pure cryptographic overhead. 
To isolate this effect, we ran the TLS benchmark with session resumption enabled.
With resumption, native and container throughput rise substantially for the fast combinations (Dilithium2+Kyber512: \SI{1.6}{\times} and \SI{1.5}{\times} respectively).
The unikernel stays relatively consistent (\SI{1.1}{\times}), indicating that it is limited by communication and not cryptographic computation.
Slower operations using e.g. SPHINCS+ still change drastically ($\approx\SI{500}{\times}$), showing the impact of the cryptography in these cases. 
However, this difference in network stacks would also affect real-world applications, and, moreover, the primitive benchmarks are unaffected.

% \paragraph{Single platform, single campaign.}
All results stem from one \gls{pi4} and from a single \SI{20}{\second} per primitive and \SI{30}{\second} TLS operation measurement window.
Within-window standard deviations are recorded for primitives, but there are no independent repetitions from which cross-run confidence intervals could be derived.
%
%\paragraph{Primitive and TLS results are from different campaigns.}
The primitive dataset was collected with a more mature toolchain, whereas the TLS speed and TLS power datasets stem from an earlier campaign.
However, most changes impact robustness, accuracy, and context, e.g., switching from the FNB58 USB meter to the HMC8043 power supply.
Thus, we can use the primitive runs as cross-comparisons between primitive-level and TLS-level results to verify and analyze qualitative findings observed inside the TLS run.
For quantitative analysis, however, the primitive results should be favored.

% \paragraph{Memory metric granularity.}
For the primitive campaign, memory is sampled by polling \texttt{proc} information.
Which process gets tracked is a choice, and for the container and the unikernel there are multiple candidates that could each be measured.
For the container, the whole Docker/containerd daemon, the \texttt{containerd-shim} process, and the actual containerized benchmark process are three differently sized quantities.
To still include some overhead from the container environment we decided to measure the shim's footprint at the cost of washing out more of the cryptographic algorithm's cost.
For the unikernel, the whole host-side \texttt{qemu-system-aarch64} process and the guest-internal heap footprint are likewise potential quantities.
The driver tracks the former, against a fixed 1\,GiB guest allocation for the primitive benchmarks.
Peak usage and allocation behavior inside either the container or the unikernel guest are therefore not visible, and the container figures in particular should not be read as purely reflecting the benchmarked algorithms. 
Future work should extend the analysis onto all potential measurands to gain a more differentiated picture.
% \todo{Measure all three per-environment memory targets (whole
% process/daemon, shim, and the actual benchmarked/guest-heap process)
% instead of a single fixed choice.}

% \paragraph{Power measurement.}
Both instruments used for power measurements measure whole-system input power to understand the power draw of the whole system during cryptographic operations. 
The primitive campaign uses a lab-grade programmable supply (HMC8043) with trace alignment, which improves on the earlier TLS setup.
The HMC8043 trace additionally records an idle baseline immediately before each primitive operation, which drifts by up to $\approx$600\,mW between measurement windows independent of which environment is running.
This drift is large relative to many operations' own power draw above idle, so we report energy with this baseline subtracted rather than raw energy.
The TLS campaign's FNB58 trace does not record a comparable per-handshake baseline, so its energy-per-handshake figures are not baseline-subtracted.
This has to be considered when deriving conclusions from the results.

Finally, we measure performance, memory, and power only. We do not evaluate isolation strength, side channels, boot time, image size, or build/toolchain effort.
Additionally, our port is a research prototype, which could be further tuned. 
However, we decided against it as its setup already required significantly higher effort than the container version. 
Still, its performance may not represent a tuned production unikernel. 

% Conclusion. Restates only findings established in section 6/7; future work
% items map to repository TODOs (T9, T10, T11) and the repo's own task list.
\section{Conclusion}
\label{sec:conclusion}

We presented the first measurement study of post-quantum cryptography under lightweight virtualization on embedded hardware, comparing native execution, Docker containers, and Unikraft unikernels with an identical OpenSSL/liboqs stack on a \gls{pi4}.
We find that containers are essentially free for post-quantum computation, whereas unikernel overhead is strongly algorithm-dependent.
For ECDSA, RSA, CROSS, FrodoKEM, Classic McEliece, and, notably, the NIST-standardized ML-KEM, ML-DSA, and the SHAKE variants of SPHINCS+, the overhead is near-native, while it concentrates in the alternates ($1.28$--$1.53\times$ for BIKE, HQC, and MAYO) and, above all, in Falcon signing ($17.8$--$19.2\times$).
The SHA-2 variants of SPHINCS+ run faster than native ($0.54$--$0.67\times$) in our environment.
Separately, we identify a build artifact: the pre-standardization Kyber and Dilithium implementations run $1.57$--$2.22\times$ slower in the unikernel than the ML-KEM and ML-DSA that superseded them, even though the same AArch64-optimized backend is compiled into both, evidently going unused at runtime in the unikernel guest.
Per-operation energy tracks execution time in every environment. In TLS, both environments cost 14--27\% of native handshake throughput while the cryptography is cheap, and nothing once it is expensive.
System power differs by at most a few percent throughout.
This shows that algorithm selection matters most, but the careful construction and verification of which implementation and cryptographic libraries are used can have a significant impact as well.

Future work includes root-causing the Falcon signing anomaly, verifying that the  SPHINCS+ SHA-2 outlier is caused by the linked cryptographic libraries, isolating the entropy-handling effect experimentally, and adding boot-time and image-size measurements to complete the deployment trade-off.
The comparison could also be broadened by also including  MicroVMs.
Furthermore, migrating the current implementation to Unikraft v0.21.0 is also on the list of future work. 
Version 0.21.0 introduces support for reseeding the Cryptographically Secure Pseudo-Random Number Generator \cite{unikraft.release.0.21.0}, which would be of interest for evaluating the entropy-handling. 
A dedicated evaluation of Unikraft's native CSPRNG in terms of security and performance might be relevant independent of its impact on \gls{pqc}.

%% ------------------------------------------------------------- bibliography
\bibliographystyle{ACM-Reference-Format}
\bibliography{pqc-vs-unikernel}

%% ------------------------------------------------------------- appendices
\appendix

% Ethics and reproducibility. AsiaCCS requires a formal Open Science appendix
% and (if applicable) an Ethical Considerations appendix after the references.

% \section{Open Science}
% \label{app:openscience}
% The complete artifacts and results are available at: \todo{add zenodo link, when paper accepted}.
% Fixes have been contributed to the respective projects. \todo{Bring liboqs fixes upstream, when paper accepted}
% % the benchmarking framework (build scripts for the native, container,
% % and unikernel environments, including Unikraft patches), the orchestration
% % and analysis scripts, and the raw JSON result files underlying every table.
% % We will make all of these available.
% % \todo{Insert anonymized artifact link (e.g., anonymous.4open.science) before
% % submission; the artifact currently resides on an institution-internal GitLab
% % instance (T2).}

\section{Ethical Considerations}
\label{app:ethics}
As far as we are aware, this work does not raise ethical concerns.
% This measurement study involves no human subjects, personal data, production
% systems, or previously unknown vulnerabilities; all experiments ran on
% author-owned hardware against author-operated endpoints. We therefore
% believe this work raises no ethical concerns. This appendix is included in
% the spirit of the CFP's guidance to err on the side of disclosure.
% This work benchmarks publicly available cryptographic implementations on
% hardware owned by the authors. No human subjects, personal data, production
% systems, or third-party infrastructure are involved, and no vulnerabilities
% are disclosed. We do not identify ethical concerns beyond the general
% observation that performance results should not be misread as security
% assessments (Section~\ref{sec:limitations}).

\section{Use of AI}
In this work AI has been used for:
\begin{itemize}
    \item Coding assistance for creating the benchmark and analysis scripts.
    \item Drafting, proofreading, grammar and spelling correction of the article itself.
\end{itemize}
The authors remain solely responsible for the content and results of this work.

\section{Full Primitive Results}
\label{app:fulltables}
Tables~\ref{tab:full-sig} and~\ref{tab:full-kem} list the mean per-operation
times for all 60~benchmarked algorithm configurations.
% Auto-generated from the raw JSON by paper/scripts/make_tables.py.
%%% AUTO-GENERATED by paper/scripts/make_tables.py from
%%% results_speed_2026-08-12_21-05-00.json -- DO NOT EDIT BY HAND.
%%% Ntv. column: native value (mean_us); Cnt./Uk. columns: value with ratio to native in
%%% parentheses.
\begin{table*}[p]
\centering
\caption{Full signature-primitive results. Mean time per operation in \unit{\micro\second} over a \SI{20}{\second} window per operation. Ntv.\ is the native value; Cnt.\ and Uk.\ give the container/unikernel value, with the ratio to native in parentheses.}
\label{tab:full-sig}
\footnotesize
\setlength{\tabcolsep}{3pt}
\begin{tabular}{l | rrr|rrr|rrr}
\toprule
 & \multicolumn{3}{c|}{Keygen} & \multicolumn{3}{c|}{Sign} & \multicolumn{3}{c}{Verify} \\
Algorithm & Ntv. & Cnt. & Uk. & Ntv. & Cnt. & Uk. & Ntv. & Cnt. & Uk. \\
\midrule
Dilithium2 & 158 & 158 (1.00) & 208 (1.32) & 467 & 465 (1.00) & 938 (2.01) & 152 & 152 (1.00) & 225 (1.48) \\
Dilithium3 & 293 & 294 (1.00) & 371 (1.27) & 736 & 738 (1.00) & 1526 (2.07) & 253 & 254 (1.00) & 362 (1.43) \\
Dilithium5 & 449 & 450 (1.00) & 563 (1.25) & 950 & 937 (0.99) & 1824 (1.92) & 435 & 435 (1.00) & 591 (1.36) \\
ML-DSA-44 & 208 & 209 (1.00) & 209 (1.00) & 915 & 926 (1.01) & 933 (1.02) & 223 & 224 (1.00) & 226 (1.01) \\
ML-DSA-65 & 369 & 370 (1.00) & 369 (1.00) & 1492 & 1513 (1.01) & 1518 (1.02) & 360 & 360 (1.00) & 363 (1.01) \\
ML-DSA-87 & 566 & 565 (1.00) & 562 (0.99) & 1817 & 1823 (1.00) & 1816 (1.00) & 588 & 588 (1.00) & 592 (1.01) \\
Falcon-512 & 17238 & 17771 (1.03) & 40995 (2.38) & 633 & 629 (0.99) & 11234 (17.75) & 90 & 90 (1.00) & 128 (1.41) \\
Falcon-1024 & 50310 & 48501 (0.96) & 117906 (2.34) & 1285 & 1280 (1.00) & 24506 (19.07) & 177 & 177 (1.00) & 255 (1.44) \\
Falcon-padded-512 & 17639 & 17788 (1.01) & 40056 (2.27) & 627 & 628 (1.00) & 11230 (17.90) & 91 & 90 (1.00) & 127 (1.40) \\
Falcon-padded-1024 & 49660 & 48761 (0.98) & 110284 (2.22) & 1277 & 1280 (1.00) & 24515 (19.20) & 177 & 177 (1.00) & 255 (1.45) \\
SPHINCS+-SHA2-128f-simple & 5367 & 5201 (0.97) & 3161 (0.59) & 125513 & 123323 (0.98) & 73635 (0.59) & 7239 & 7329 (1.01) & 4449 (0.61) \\
SPHINCS+-SHA2-128s-simple & 345296 & 334043 (0.97) & 202330 (0.59) & 2629679 & 2545501 (0.97) & 1530117 (0.58) & 2641 & 2421 (0.92) & 1590 (0.60) \\
SPHINCS+-SHA2-192f-simple & 7937 & 7705 (0.97) & 4987 (0.63) & 209350 & 202809 (0.97) & 128942 (0.62) & 11245 & 11002 (0.98) & 6888 (0.61) \\
SPHINCS+-SHA2-192s-simple & 510142 & 489995 (0.96) & 310677 (0.61) & 4698340 & 4513138 (0.96) & 2865978 (0.61) & 3829 & 3849 (1.01) & 2567 (0.67) \\
SPHINCS+-SHA2-256f-simple & 22561 & 20186 (0.89) & 12742 (0.56) & 462626 & 419039 (0.91) & 261087 (0.56) & 11943 & 11217 (0.94) & 6928 (0.58) \\
SPHINCS+-SHA2-256s-simple & 375745 & 338981 (0.90) & 203649 (0.54) & 4637549 & 4221643 (0.91) & 2522469 (0.54) & 6280 & 5792 (0.92) & 3431 (0.55) \\
SPHINCS+-SHAKE-128f-simple & 5440 & 5465 (1.00) & 5445 (1.00) & 127668 & 128252 (1.00) & 125275 (0.98) & 7427 & 7504 (1.01) & 7324 (0.99) \\
SPHINCS+-SHAKE-128s-simple & 352488 & 353836 (1.00) & 333623 (0.95) & 2677162 & 2681182 (1.00) & 2535920 (0.95) & 2591 & 2600 (1.00) & 2356 (0.91) \\
SPHINCS+-SHAKE-192f-simple & 8011 & 8003 (1.00) & 7768 (0.97) & 208003 & 208158 (1.00) & 200594 (0.96) & 10983 & 11012 (1.00) & 11047 (1.01) \\
SPHINCS+-SHAKE-192s-simple & 516730 & 517272 (1.00) & 492093 (0.95) & 4645198 & 4650068 (1.00) & 4425673 (0.95) & 3871 & 3846 (0.99) & 3666 (0.95) \\
SPHINCS+-SHAKE-256f-simple & 22132 & 22187 (1.00) & 20826 (0.94) & 446433 & 448006 (1.00) & 421752 (0.94) & 11625 & 11822 (1.02) & 11292 (0.97) \\
SPHINCS+-SHAKE-256s-simple & 328001 & 328805 (1.00) & 335812 (1.02) & 4039135 & 4044271 (1.00) & 4038059 (1.00) & 5803 & 5851 (1.01) & 5487 (0.95) \\
MAYO-1 & 1070 & 1071 (1.00) & 1709 (1.60) & 2381 & 2377 (1.00) & 2852 (1.20) & 729 & 743 (1.02) & 1459 (2.00) \\
MAYO-2 & 2374 & 2368 (1.00) & 3061 (1.29) & 3327 & 3324 (1.00) & 3929 (1.18) & 792 & 795 (1.00) & 1775 (2.24) \\
MAYO-3 & 3953 & 3951 (1.00) & 6113 (1.55) & 8894 & 8914 (1.00) & 10555 (1.19) & 2601 & 2590 (1.00) & 5082 (1.95) \\
MAYO-5 & 10335 & 10341 (1.00) & 15722 (1.52) & 23076 & 23065 (1.00) & 27522 (1.19) & 6564 & 6374 (0.97) & 12324 (1.88) \\
cross-rsdp-128-balanced & 57 & 57 (1.00) & 56 (0.97) & 3331 & 3306 (0.99) & 3477 (1.04) & 1931 & 1928 (1.00) & 1987 (1.03) \\
cross-rsdp-128-fast & 57 & 57 (1.00) & 56 (0.98) & 1821 & 1831 (1.01) & 1903 (1.04) & 1037 & 1034 (1.00) & 1072 (1.03) \\
cross-rsdp-128-small & 57 & 57 (1.01) & 56 (0.98) & 12435 & 12429 (1.00) & 13056 (1.05) & 7407 & 7388 (1.00) & 7531 (1.02) \\
cross-rsdp-192-balanced & 121 & 122 (1.00) & 119 (0.98) & 7864 & 7873 (1.00) & 7862 (1.00) & 4307 & 4293 (1.00) & 4201 (0.98) \\
cross-rsdp-192-fast & 121 & 121 (1.00) & 118 (0.98) & 4443 & 4426 (1.00) & 4464 (1.00) & 2566 & 2516 (0.98) & 2527 (0.98) \\
cross-rsdp-256-balanced & 212 & 213 (1.00) & 205 (0.97) & 15213 & 15179 (1.00) & 15972 (1.05) & 7338 & 7334 (1.00) & 7301 (0.99) \\
cross-rsdp-256-fast & 213 & 212 (1.00) & 205 (0.96) & 9353 & 9312 (1.00) & 9769 (1.04) & 5200 & 5185 (1.00) & 5494 (1.06) \\
cross-rsdpg-128-balanced & 29 & 29 (1.01) & 29 (1.01) & 2541 & 2551 (1.00) & 2591 (1.02) & 1545 & 1562 (1.01) & 1607 (1.04) \\
cross-rsdpg-128-fast & 29 & 29 (1.01) & 29 (1.00) & 1297 & 1297 (1.00) & 1353 (1.04) & 787 & 788 (1.00) & 810 (1.03) \\
cross-rsdpg-128-small & 29 & 29 (1.01) & 29 (1.01) & 9098 & 9135 (1.00) & 9290 (1.02) & 5511 & 5519 (1.00) & 5703 (1.03) \\
cross-rsdpg-192-balanced & 53 & 53 (1.00) & 56 (1.06) & 3808 & 3816 (1.00) & 3834 (1.01) & 2340 & 2347 (1.00) & 2340 (1.00) \\
cross-rsdpg-192-fast & 53 & 53 (1.00) & 56 (1.05) & 3002 & 2976 (0.99) & 2963 (0.99) & 1865 & 1864 (1.00) & 1865 (1.00) \\
cross-rsdpg-192-small & 53 & 53 (1.01) & 56 (1.06) & 14005 & 14057 (1.00) & 13921 (0.99) & 8796 & 8765 (1.00) & 8734 (0.99) \\
cross-rsdpg-256-balanced & 81 & 82 (1.01) & 88 (1.08) & 6545 & 6500 (0.99) & 6636 (1.01) & 3836 & 3830 (1.00) & 3969 (1.03) \\
cross-rsdpg-256-fast & 81 & 81 (1.01) & 88 (1.09) & 5033 & 5027 (1.00) & 5237 (1.04) & 3129 & 3123 (1.00) & 3219 (1.03) \\
\hdashline
ECDSA & 43 & 43 (1.01) & 41 (0.96) & 97 & 97 (1.00) & 96 (0.99) & 291 & 290 (1.00) & 290 (0.99) \\
RSA-2048 & 319292 & 368907 (1.16) & 355618 (1.11) & 4967 & 4980 (1.00) & 4980 (1.00) & 133 & 134 (1.01) & 134 (1.01) \\
\bottomrule
\end{tabular}
\end{table*}

%%% AUTO-GENERATED by paper/scripts/make_tables.py from
%%% results_speed_2026-08-12_21-05-00.json -- DO NOT EDIT BY HAND.
%%% Ntv. column: native value (mean_us); Cnt./Uk. columns: value with ratio to native in
%%% parentheses.
\begin{table*}[p]
\centering
\caption{Full KEM-primitive results (Structure as in Table~\ref{tab:full-sig}).}
\label{tab:full-kem}
\footnotesize
\setlength{\tabcolsep}{3pt}
\begin{tabular}{l | rrr|rrr|rrr}
\toprule
 & \multicolumn{3}{c|}{Keygen} & \multicolumn{3}{c|}{Encaps} & \multicolumn{3}{c}{Decaps} \\
Algorithm & Ntv. & Cnt. & Uk. & Ntv. & Cnt. & Uk. & Ntv. & Cnt. & Uk. \\
\midrule
BIKE-L1 & 42924 & 42878 (1.00) & 54991 (1.28) & 2177 & 2178 (1.00) & 2810 (1.29) & 35255 & 35090 (1.00) & 43744 (1.24) \\
BIKE-L3 & 133308 & 133185 (1.00) & 170274 (1.28) & 6682 & 6683 (1.00) & 8605 (1.29) & 110868 & 110283 (0.99) & 136565 (1.23) \\
BIKE-L5 & 334433 & 333915 (1.00) & 427914 (1.28) & 16743 & 16756 (1.00) & 21663 (1.29) & 276644 & 276171 (1.00) & 343380 (1.24) \\
Classic-McEliece-348864 & 344534 & 357498 (1.04) & 364574 (1.06) & 169 & 169 (1.00) & 254 (1.51) & 56628 & 56681 (1.00) & 55146 (0.97) \\
Classic-McEliece-348864f & 160245 & 158878 (0.99) & 154329 (0.96) & 168 & 167 (0.99) & 254 (1.51) & 56650 & 56647 (1.00) & 55141 (0.97) \\
Classic-McEliece-460896 & 1389270 & 1209027 (0.87) & 1288431 (0.93) & 350 & 354 (1.01) & 571 (1.63) & 84538 & 84569 (1.00) & 88917 (1.05) \\
Classic-McEliece-460896f & 536332 & 547725 (1.02) & 486712 (0.91) & 343 & 353 (1.03) & 568 (1.65) & 84500 & 84554 (1.00) & 88946 (1.05) \\
HQC-128 & 5216 & 5204 (1.00) & 6732 (1.29) & 10496 & 10513 (1.00) & 13422 (1.28) & 16640 & 16597 (1.00) & 20296 (1.22) \\
HQC-192 & 15877 & 15879 (1.00) & 20509 (1.29) & 31860 & 31854 (1.00) & 40983 (1.29) & 49512 & 48832 (0.99) & 61736 (1.25) \\
HQC-256 & 29152 & 29128 (1.00) & 37499 (1.29) & 58484 & 58570 (1.00) & 74982 (1.28) & 90338 & 90515 (1.00) & 112967 (1.25) \\
Kyber512 & 46 & 47 (1.02) & 86 (1.86) & 52 & 52 (1.00) & 104 (1.98) & 43 & 43 (1.00) & 117 (2.73) \\
Kyber768 & 68 & 68 (1.00) & 136 (2.01) & 79 & 79 (1.01) & 164 (2.08) & 69 & 69 (1.00) & 182 (2.63) \\
Kyber1024 & 100 & 100 (1.01) & 205 (2.06) & 115 & 115 (1.01) & 239 (2.08) & 104 & 103 (0.99) & 265 (2.54) \\
ML-KEM-512 & 79 & 79 (1.01) & 86 (1.09) & 94 & 95 (1.01) & 96 (1.02) & 115 & 114 (1.00) & 117 (1.02) \\
ML-KEM-768 & 130 & 131 (1.01) & 137 (1.06) & 151 & 150 (1.00) & 151 (1.00) & 179 & 179 (1.00) & 181 (1.01) \\
ML-KEM-1024 & 200 & 200 (1.00) & 205 (1.03) & 286 & 223 (0.78) & 225 (0.79) & 262 & 261 (1.00) & 262 (1.00) \\
FrodoKEM-640-AES & 17011 & 16934 (1.00) & 15680 (0.92) & 17227 & 17253 (1.00) & 15908 (0.92) & 17300 & 17212 (0.99) & 15758 (0.91) \\
FrodoKEM-640-SHAKE & 5810 & 5839 (1.00) & 5759 (0.99) & 6573 & 6539 (0.99) & 6508 (0.99) & 6483 & 6490 (1.00) & 6453 (1.00) \\
FrodoKEM-976-AES & 39204 & 39142 (1.00) & 35957 (0.92) & 39906 & 39811 (1.00) & 36642 (0.92) & 39704 & 39703 (1.00) & 36204 (0.91) \\
FrodoKEM-976-SHAKE & 13127 & 13042 (0.99) & 12941 (0.99) & 14527 & 14452 (0.99) & 14477 (1.00) & 14389 & 14495 (1.01) & 14358 (1.00) \\
FrodoKEM-1344-AES & 74155 & 74077 (1.00) & 67467 (0.91) & 75064 & 75293 (1.00) & 69026 (0.92) & 75071 & 75245 (1.00) & 68942 (0.92) \\
FrodoKEM-1344-SHAKE & 23771 & 23699 (1.00) & 23399 (0.98) & 26549 & 26386 (0.99) & 26352 (0.99) & 26365 & 26198 (0.99) & 26317 (1.00) \\
\hdashline
ECDHE & 62 & 62 (0.99) & 62 (1.00) & 245 & 246 (1.00) & 247 (1.01) & --- & --- & --- \\
X25519 & 83 & 84 (1.00) & 73 (0.88) & 249 & 249 (1.00) & 240 (0.96) & 248 & 248 (1.00) & 240 (0.96) \\
\hdashline[1pt/1pt]
x25519\_Kyber512 & 132 & 133 (1.00) & 163 (1.23) & 304 & 304 (1.00) & 346 (1.14) & 295 & 295 (1.00) & 360 (1.22) \\
x25519\_Kyber768 & 153 & 154 (1.00) & 214 (1.39) & 331 & 332 (1.00) & 405 (1.22) & 320 & 321 (1.00) & 425 (1.33) \\
x25519\_Kyber1024 & 185 & 186 (1.00) & 282 (1.52) & 367 & 367 (1.00) & 479 (1.31) & 356 & 355 (1.00) & 507 (1.42) \\
\bottomrule
\end{tabular}
\end{table*}

Tables~\ref{tab:full-mem-sig} and~\ref{tab:full-mem-kem} list the mean
per-operation resident memory for the same 60~configurations.
% Auto-generated from the raw JSON by paper/scripts/make_tables.py.
%%% AUTO-GENERATED by paper/scripts/make_tables.py from
%%% results_memory_2026-08-13_12-26-35.json -- DO NOT EDIT BY HAND.
%%% Ntv. column: native value (memory_rss_avg_kb); Cnt./Uk. columns: value.
\begin{table*}[p]
\centering
\caption{Full signature-primitive resident memory results. Mean per-operation resident memory in \unit{\mebi\byte} for the native, container, and unikernel environments.}
\label{tab:full-mem-sig}
\footnotesize
\setlength{\tabcolsep}{3pt}
\begin{tabular}{l | rrr|rrr|rrr}
\toprule
 & \multicolumn{3}{c|}{Keygen} & \multicolumn{3}{c|}{Sign} & \multicolumn{3}{c}{Verify} \\
Algorithm & Ntv. & Cnt. & Uk. & Ntv. & Cnt. & Uk. & Ntv. & Cnt. & Uk. \\
\midrule
Dilithium2 & 4.3 & 13.6 & 57.5 & 4.3 & 13.6 & 57.5 & 4.3 & 13.8 & 57.5 \\
Dilithium3 & 4.3 & 13.6 & 57.4 & 4.3 & 13.7 & 57.5 & 4.3 & 13.6 & 57.5 \\
Dilithium5 & 4.3 & 13.8 & 57.6 & 4.3 & 13.6 & 57.5 & 4.3 & 13.7 & 57.6 \\
ML-DSA-44 & 4.1 & 13.8 & 57.5 & 4.1 & 13.6 & 57.5 & 4.1 & 13.7 & 57.4 \\
ML-DSA-65 & 4.1 & 13.6 & 57.4 & 4.1 & 14.1 & 57.4 & 4.1 & 13.7 & 57.4 \\
ML-DSA-87 & 4.1 & 13.6 & 57.6 & 4.1 & 13.6 & 57.6 & 4.1 & 13.6 & 57.6 \\
Falcon-512 & 4.1 & 13.6 & 57.5 & 4.1 & 13.6 & 57.4 & 4.1 & 13.8 & 57.5 \\
Falcon-1024 & 4.1 & 13.7 & 57.6 & 4.1 & 13.8 & 57.5 & 4.1 & 13.8 & 57.4 \\
Falcon-padded-512 & 4.0 & 13.8 & 57.5 & 4.0 & 13.8 & 57.5 & 4.0 & 13.9 & 57.5 \\
Falcon-padded-1024 & 4.3 & 13.6 & 57.5 & 4.3 & 13.6 & 57.5 & 4.3 & 13.8 & 57.5 \\
SPHINCS+-SHA2-128f-simple & 4.1 & 13.6 & 57.5 & 4.1 & 13.5 & 57.6 & 4.1 & 13.6 & 57.5 \\
SPHINCS+-SHA2-128s-simple & 4.0 & 13.5 & 57.4 & 4.0 & 13.8 & 57.8 & 4.0 & 13.8 & 57.4 \\
SPHINCS+-SHA2-192f-simple & 4.1 & 13.5 & 57.6 & 4.1 & 13.6 & 57.5 & 4.1 & 13.6 & 57.5 \\
SPHINCS+-SHA2-192s-simple & 4.1 & 13.9 & 57.5 & 4.1 & 14.0 & 57.8 & 4.1 & 13.9 & 57.5 \\
SPHINCS+-SHA2-256f-simple & 4.1 & 13.8 & 57.5 & 4.1 & 13.6 & 57.6 & 4.1 & 13.7 & 57.5 \\
SPHINCS+-SHA2-256s-simple & 4.1 & 13.7 & 57.5 & 4.1 & 13.8 & 57.7 & 4.1 & 13.7 & 57.6 \\
SPHINCS+-SHAKE-128f-simple & 3.9 & 13.6 & 57.5 & 3.9 & 13.8 & 57.5 & 3.9 & 13.7 & 57.5 \\
SPHINCS+-SHAKE-128s-simple & 3.9 & 13.7 & 57.5 & 3.9 & 13.7 & 57.6 & 3.9 & 13.7 & 57.6 \\
SPHINCS+-SHAKE-192f-simple & 3.9 & 13.9 & 57.5 & 3.9 & 13.7 & 57.5 & 3.9 & 13.8 & 57.5 \\
SPHINCS+-SHAKE-192s-simple & 3.9 & 13.5 & 57.5 & 3.9 & 13.7 & 57.9 & 3.9 & 13.6 & 57.5 \\
SPHINCS+-SHAKE-256f-simple & 4.0 & 13.9 & 57.5 & 4.0 & 13.8 & 57.6 & 4.0 & 13.5 & 57.5 \\
SPHINCS+-SHAKE-256s-simple & 3.9 & 13.7 & 57.6 & 3.9 & 13.9 & 57.7 & 3.9 & 13.8 & 57.6 \\
MAYO-1 & 4.5 & 13.7 & 57.8 & 4.5 & 13.6 & 57.8 & 4.5 & 13.8 & 57.7 \\
MAYO-2 & 4.4 & 13.5 & 57.7 & 4.4 & 13.8 & 57.7 & 4.3 & 13.6 & 57.6 \\
MAYO-3 & 4.4 & 13.6 & 57.8 & 4.4 & 13.7 & 57.8 & 4.4 & 13.6 & 57.8 \\
MAYO-5 & 4.9 & 13.8 & 58.3 & 4.9 & 13.6 & 58.2 & 4.9 & 13.8 & 58.3 \\
cross-rsdp-128-balanced & 4.3 & 13.7 & 57.6 & 4.3 & 13.9 & 57.6 & 4.3 & 13.6 & 57.6 \\
cross-rsdp-128-fast & 4.0 & 13.7 & 57.5 & 4.0 & 13.5 & 57.5 & 4.0 & 13.5 & 57.5 \\
cross-rsdp-128-small & 4.8 & 13.6 & 58.2 & 4.8 & 13.8 & 58.1 & 4.8 & 13.9 & 58.2 \\
cross-rsdp-192-balanced & 4.4 & 13.8 & 57.9 & 4.4 & 13.7 & 57.8 & 4.4 & 13.8 & 57.8 \\
cross-rsdp-192-fast & 4.1 & 13.6 & 57.8 & 4.1 & 13.8 & 57.7 & 4.1 & 13.7 & 57.7 \\
cross-rsdp-256-balanced & 4.8 & 13.9 & 58.2 & 4.8 & 13.8 & 58.2 & 4.8 & 13.8 & 58.2 \\
cross-rsdp-256-fast & 4.3 & 13.9 & 57.9 & 4.3 & 13.7 & 57.8 & 4.3 & 13.6 & 57.8 \\
cross-rsdpg-128-balanced & 4.0 & 13.8 & 57.5 & 4.0 & 13.8 & 57.5 & 4.0 & 13.6 & 57.5 \\
cross-rsdpg-128-fast & 4.0 & 13.8 & 57.5 & 4.0 & 13.6 & 57.6 & 4.0 & 13.6 & 57.4 \\
cross-rsdpg-128-small & 4.5 & 13.8 & 57.9 & 4.5 & 13.8 & 57.8 & 4.5 & 13.8 & 57.8 \\
cross-rsdpg-192-balanced & 4.1 & 13.7 & 57.6 & 4.1 & 13.9 & 57.6 & 4.1 & 13.9 & 57.6 \\
cross-rsdpg-192-fast & 4.1 & 13.4 & 57.6 & 4.1 & 13.8 & 57.7 & 4.1 & 13.5 & 57.8 \\
cross-rsdpg-192-small & 4.6 & 13.9 & 58.3 & 4.6 & 13.6 & 58.3 & 4.6 & 13.8 & 58.2 \\
cross-rsdpg-256-balanced & 4.5 & 13.4 & 57.8 & 4.5 & 13.7 & 57.8 & 4.5 & 13.7 & 57.8 \\
cross-rsdpg-256-fast & 4.3 & 13.9 & 57.7 & 4.3 & 13.9 & 57.7 & 4.3 & 13.5 & 57.7 \\
\hdashline
ECDSA & 4.0 & 13.7 & 76.1 & 4.0 & 13.8 & 76.1 & 4.0 & 13.6 & 76.1 \\
RSA-2048 & 3.9 & 13.5 & 78.8 & 3.9 & 13.5 & 78.3 & 3.9 & 13.8 & 78.4 \\
\bottomrule
\end{tabular}
\end{table*}

%%% AUTO-GENERATED by paper/scripts/make_tables.py from
%%% results_memory_2026-08-13_12-26-35.json -- DO NOT EDIT BY HAND.
%%% Ntv. column: native value (memory_rss_avg_kb); Cnt./Uk. columns: value.
\begin{table*}[p]
\centering
\caption{Full KEM-primitive resident memory results. Mean per-operation resident memory in \unit{\mebi\byte} for the native, container, and unikernel environments.}
\label{tab:full-mem-kem}
\footnotesize
\setlength{\tabcolsep}{3pt}
\begin{tabular}{l | rrr|rrr|rrr}
\toprule
 & \multicolumn{3}{c|}{Keygen} & \multicolumn{3}{c|}{Encaps} & \multicolumn{3}{c}{Decaps} \\
Algorithm & Ntv. & Cnt. & Uk. & Ntv. & Cnt. & Uk. & Ntv. & Cnt. & Uk. \\
\midrule
BIKE-L1 & 3.9 & 13.5 & 57.4 & 3.9 & 13.8 & 57.4 & 4.0 & 13.7 & 57.4 \\
BIKE-L3 & 4.0 & 13.5 & 57.4 & 4.0 & 13.2 & 57.4 & 4.0 & 13.9 & 57.5 \\
BIKE-L5 & 4.0 & 13.8 & 57.7 & 4.0 & 13.6 & 57.6 & 4.3 & 13.2 & 57.8 \\
Classic-McEliece-348864 & 4.5 & 13.6 & 58.1 & 4.5 & 13.5 & 58.0 & 4.5 & 13.8 & 58.0 \\
Classic-McEliece-348864f & 4.5 & 14.0 & 58.1 & 4.5 & 13.9 & 58.0 & 4.5 & 13.7 & 58.0 \\
Classic-McEliece-460896 & 5.3 & 13.9 & 59.0 & 5.2 & 13.9 & 58.7 & 5.2 & 13.9 & 58.7 \\
Classic-McEliece-460896f & 5.3 & 13.9 & 58.9 & 5.2 & 13.6 & 58.7 & 5.2 & 13.7 & 58.7 \\
HQC-128 & 4.0 & 13.6 & 57.4 & 4.1 & 13.8 & 57.4 & 4.1 & 13.7 & 57.6 \\
HQC-192 & 4.0 & 13.8 & 57.4 & 4.1 & 13.8 & 57.5 & 4.1 & 13.8 & 57.5 \\
HQC-256 & 4.0 & 13.9 & 57.5 & 4.3 & 13.8 & 57.6 & 4.3 & 13.6 & 57.7 \\
Kyber512 & 4.1 & 13.8 & 57.4 & 4.1 & 13.7 & 57.4 & 4.1 & 13.9 & 57.4 \\
Kyber768 & 4.0 & 13.8 & 57.4 & 4.0 & 14.1 & 57.5 & 4.0 & 13.8 & 57.5 \\
Kyber1024 & 4.0 & 13.6 & 57.4 & 4.0 & 13.8 & 57.4 & 4.0 & 13.5 & 57.4 \\
ML-KEM-512 & 4.0 & 13.7 & 57.4 & 4.0 & 13.6 & 57.4 & 4.0 & 13.7 & 57.4 \\
ML-KEM-768 & 4.0 & 14.0 & 57.4 & 4.0 & 13.7 & 57.4 & 4.0 & 13.9 & 57.4 \\
ML-KEM-1024 & 4.0 & 13.6 & 57.5 & 4.0 & 13.9 & 57.4 & 4.0 & 13.6 & 57.4 \\
FrodoKEM-640-AES & 3.9 & 13.9 & 57.5 & 3.9 & 13.7 & 57.5 & 3.9 & 13.9 & 57.5 \\
FrodoKEM-640-SHAKE & 3.9 & 13.9 & 57.5 & 3.9 & 13.9 & 57.5 & 3.9 & 13.4 & 57.5 \\
FrodoKEM-976-AES & 3.9 & 13.5 & 57.6 & 4.0 & 13.6 & 57.5 & 4.0 & 13.7 & 57.6 \\
FrodoKEM-976-SHAKE & 4.0 & 13.6 & 57.5 & 4.0 & 13.5 & 57.5 & 4.0 & 13.7 & 57.7 \\
FrodoKEM-1344-AES & 4.1 & 13.7 & 57.5 & 4.1 & 13.8 & 57.6 & 4.3 & 13.6 & 57.6 \\
FrodoKEM-1344-SHAKE & 4.0 & 13.6 & 57.5 & 4.1 & 13.9 & 57.6 & 4.1 & 13.8 & 57.6 \\
\hdashline
ECDHE & 3.9 & 13.5 & 75.9 & 4.1 & 13.8 & 77.0 & --- & --- & --- \\
X25519 & 3.8 & 13.8 & 75.7 & 3.8 & 13.6 & 75.9 & 3.8 & 13.6 & 75.7 \\
\hdashline[1pt/1pt]
x25519\_Kyber512 & 4.3 & 13.6 & 75.6 & 4.4 & 13.8 & 75.8 & 4.4 & 13.8 & 75.7 \\
x25519\_Kyber768 & 4.3 & 13.7 & 75.5 & 4.3 & 13.5 & 75.7 & 4.3 & 13.5 & 75.8 \\
x25519\_Kyber1024 & 4.4 & 13.6 & 75.6 & 4.4 & 13.6 & 75.7 & 4.4 & 13.8 & 75.7 \\
\bottomrule
\end{tabular}
\end{table*}

Tables~\ref{tab:full-energy-sig} and~\ref{tab:full-energy-kem} list the mean
per-operation energy for the same 60~configurations.
% Auto-generated from the raw JSON by paper/scripts/make_tables.py.
%%% AUTO-GENERATED by paper/scripts/make_tables.py from
%%% results_power_2026-08-17_35-23.json -- DO NOT EDIT BY HAND.
%%% One row per algorithm, keygen/sign/verify as column-groups. $P$ (W) and $E$ (always mJ,
%%% field='e_marginal_per_iteration') are both baseline-subtracted (see inject_marginal_energy()).
%%% Native columns: value. Container/unikernel columns: value.
\begin{table*}[p]
\centering
\caption{Full signature-primitive power and energy results above the idle baseline. $P$ (\unit{\watt}) is the mean power draw during the operation, $E$ (\unit{\milli\joule}) is the energy per operation. Derived from the HMC8043 power trace aligned with per-operation timestamps.}
\label{tab:full-energy-sig}
\footnotesize
\setlength{\tabcolsep}{2pt}
\begin{tabular}{l | rrrrrr|rrrrrr|rrrrrr}
\toprule
 & \multicolumn{6}{c|}{Keygen} & \multicolumn{6}{c|}{Sign} & \multicolumn{6}{c}{Verify} \\
 & \multicolumn{2}{c}{Ntv.} & \multicolumn{2}{c}{Cnt.} & \multicolumn{2}{c}{Uk.} & \multicolumn{2}{c}{Ntv.} & \multicolumn{2}{c}{Cnt.} & \multicolumn{2}{c}{Uk.} & \multicolumn{2}{c}{Ntv.} & \multicolumn{2}{c}{Cnt.} & \multicolumn{2}{c}{Uk.} \\
Algorithm & P & E & P & E & P & E & P & E & P & E & P & E & P & E & P & E & P & E \\
\midrule
Dilithium2 & 0.94 & 0.149 & 1.34 & 0.222 & 1.35 & 0.284 & 1.33 & 0.616 & 1.16 & 0.558 & 1.18 & 1.11 & 1.06 & 0.161 & 1.15 & 0.180 & 1.28 & 0.291 \\
Dilithium3 & 1.32 & 0.388 & 1.35 & 0.415 & 1.07 & 0.401 & 1.34 & 0.975 & 1.34 & 1.03 & 1.39 & 2.12 & 1.36 & 0.344 & 1.36 & 0.359 & 1.02 & 0.375 \\
Dilithium5 & 1.21 & 0.542 & 1.36 & 0.644 & 1.28 & 0.729 & 1.38 & 1.30 & 1.35 & 1.32 & 1.29 & 2.41 & 1.40 & 0.612 & 1.36 & 0.618 & 1.31 & 0.785 \\
ML-DSA-44 & 1.03 & 0.214 & 1.29 & 0.310 & 1.27 & 0.268 & 1.11 & 1.28 & 1.33 & 1.28 & 1.30 & 1.24 & 1.11 & 0.249 & 1.35 & 0.315 & 1.27 & 0.292 \\
ML-DSA-65 & 1.35 & 0.500 & 1.07 & 0.413 & 1.29 & 0.484 & 1.22 & 1.82 & 1.09 & 1.68 & 1.30 & 2.02 & 1.12 & 0.405 & 1.31 & 0.493 & 1.35 & 0.495 \\
ML-DSA-87 & 1.31 & 0.742 & 1.14 & 0.678 & 1.04 & 0.587 & 1.33 & 2.39 & 1.09 & 2.19 & 1.34 & 2.50 & 1.31 & 0.771 & 1.11 & 0.749 & 1.28 & 0.767 \\
Falcon-512 & 1.12 & 19.48 & 1.12 & 20.57 & 0.89 & 36.29 & 1.22 & 0.771 & 1.20 & 0.799 & 0.84 & 9.61 & 1.18 & 0.108 & 1.02 & 0.097 & 0.91 & 0.119 \\
Falcon-1024 & 1.09 & 53.51 & 1.11 & 57.29 & 1.05 & 126 & 1.21 & 1.56 & 1.21 & 1.62 & 0.89 & 22.47 & 1.20 & 0.213 & 1.22 & 0.227 & 0.98 & 0.257 \\
Falcon-padded-512 & 1.23 & 21.39 & 1.14 & 21.10 & 1.08 & 45.06 & 0.97 & 0.612 & 1.23 & 0.812 & 1.08 & 12.46 & 1.19 & 0.109 & 1.23 & 0.115 & 1.19 & 0.155 \\
Falcon-padded-1024 & 0.83 & 40.74 & 1.22 & 62.19 & 1.05 & 130 & 1.01 & 1.30 & 1.20 & 1.63 & 1.06 & 26.74 & 1.30 & 0.233 & 1.20 & 0.222 & 1.19 & 0.313 \\
SPHINCS+-SHA2-128f-simple & 1.27 & 6.90 & 1.03 & 5.64 & 0.78 & 2.54 & 1.28 & 163 & 1.33 & 172 & 1.19 & 89.56 & 1.06 & 8.11 & 1.27 & 9.69 & 1.28 & 5.91 \\
SPHINCS+-SHA2-128s-simple & 1.27 & 513 & 1.25 & 503 & 1.19 & 283 & 1.40 & 3847 & 1.33 & 3473 & 1.23 & 1931 & 1.31 & 4.41 & 1.15 & 3.75 & 1.23 & 2.28 \\
SPHINCS+-SHA2-192f-simple & 1.31 & 10.71 & 1.34 & 11.00 & 1.27 & 6.38 & 1.39 & 298 & 1.32 & 287 & 1.27 & 169 & 1.42 & 16.35 & 1.26 & 14.70 & 1.30 & 9.08 \\
SPHINCS+-SHA2-192s-simple & 1.28 & 777 & 1.04 & 625 & 1.24 & 469 & 1.31 & 6297 & 1.31 & 6119 & 1.29 & 3772 & 1.04 & 5.88 & 1.29 & 7.63 & 1.02 & 3.37 \\
SPHINCS+-SHA2-256f-simple & 1.24 & 29.46 & 1.23 & 27.47 & 1.48 & 19.81 & 1.29 & 613 & 1.26 & 553 & 1.44 & 389 & 1.26 & 16.23 & 1.30 & 15.65 & 1.35 & 9.92 \\
SPHINCS+-SHA2-256s-simple & 1.26 & 600 & 1.27 & 559 & 1.22 & 316 & 1.32 & 6284 & 1.33 & 5750 & 1.31 & 3372 & 1.31 & 12.31 & 1.26 & 10.58 & 1.29 & 5.68 \\
SPHINCS+-SHAKE-128f-simple & 1.27 & 6.98 & 1.36 & 7.76 & 1.43 & 7.85 & 1.28 & 167 & 1.47 & 201 & 1.41 & 185 & 1.29 & 10.18 & 1.48 & 12.14 & 1.27 & 9.45 \\
SPHINCS+-SHAKE-128s-simple & 1.31 & 542 & 1.28 & 542 & 1.14 & 450 & 1.34 & 3684 & 1.34 & 3676 & 1.27 & 3302 & 1.06 & 3.62 & 1.36 & 4.54 & 1.36 & 4.43 \\
SPHINCS+-SHAKE-192f-simple & 1.10 & 9.06 & 1.27 & 10.78 & 1.25 & 10.18 & 1.22 & 260 & 1.24 & 286 & 1.26 & 261 & 1.22 & 13.93 & 1.26 & 14.66 & 1.25 & 13.89 \\
SPHINCS+-SHAKE-192s-simple & 1.19 & 735 & 1.05 & 662 & 1.31 & 787 & 1.04 & 4957 & 1.35 & 6454 & 1.32 & 6060 & 1.12 & 6.59 & 1.29 & 8.14 & 1.24 & 6.80 \\
SPHINCS+-SHAKE-256f-simple & 1.30 & 30.08 & 1.18 & 27.96 & 1.25 & 27.65 & 1.34 & 614 & 1.28 & 595 & 1.27 & 548 & 1.22 & 15.38 & 1.27 & 16.78 & 1.27 & 14.64 \\
SPHINCS+-SHAKE-256s-simple & 1.30 & 541 & 1.08 & 473 & 1.00 & 426 & 1.04 & 4296 & 1.33 & 5492 & 1.34 & 5558 & 1.17 & 9.90 & 1.39 & 11.46 & 0.95 & 7.71 \\
MAYO-1 & 1.28 & 1.38 & 1.28 & 1.43 & 0.93 & 1.61 & 1.22 & 2.90 & 1.19 & 2.96 & 1.01 & 2.94 & 1.46 & 1.08 & 1.46 & 1.12 & 1.09 & 1.61 \\
MAYO-2 & 1.39 & 3.29 & 1.22 & 3.02 & 1.19 & 3.68 & 1.18 & 3.91 & 1.16 & 4.08 & 1.18 & 4.69 & 1.45 & 1.15 & 1.42 & 1.18 & 1.03 & 1.84 \\
MAYO-3 & 1.20 & 4.73 & 1.25 & 5.17 & 1.19 & 7.35 & 1.38 & 12.30 & 1.16 & 10.89 & 1.14 & 12.27 & 1.66 & 4.27 & 1.45 & 3.91 & 1.25 & 6.50 \\
MAYO-5 & 1.25 & 12.91 & 1.41 & 15.22 & 1.18 & 18.84 & 1.09 & 25.30 & 1.37 & 33.02 & 1.18 & 32.96 & 1.69 & 10.98 & 1.44 & 9.83 & 1.25 & 15.97 \\
cross-rsdp-128-balanced & 1.20 & 0.068 & 1.36 & 0.081 & 1.32 & 0.075 & 1.28 & 4.23 & 1.49 & 5.16 & 1.25 & 4.38 & 1.24 & 2.39 & 1.42 & 2.86 & 1.20 & 2.43 \\
cross-rsdp-128-fast & 1.18 & 0.068 & 1.18 & 0.077 & 1.13 & 0.064 & 1.28 & 2.33 & 1.22 & 2.58 & 1.17 & 2.24 & 1.27 & 1.32 & 1.10 & 1.21 & 1.48 & 1.62 \\
cross-rsdp-128-small & 1.20 & 0.069 & 1.20 & 0.072 & 1.19 & 0.067 & 1.31 & 16.42 & 1.31 & 17.17 & 1.13 & 14.90 & 1.25 & 9.28 & 0.96 & 7.44 & 1.40 & 10.70 \\
cross-rsdp-192-balanced & 1.45 & 0.177 & 1.22 & 0.156 & 1.17 & 0.142 & 1.50 & 11.79 & 1.30 & 10.74 & 1.30 & 10.38 & 1.24 & 5.36 & 1.25 & 5.61 & 1.24 & 5.27 \\
cross-rsdp-192-fast & 1.04 & 0.126 & 1.39 & 0.178 & 1.18 & 0.143 & 1.51 & 6.71 & 1.29 & 5.97 & 1.27 & 5.72 & 1.50 & 3.85 & 1.30 & 3.43 & 1.26 & 3.22 \\
cross-rsdp-256-balanced & 1.21 & 0.258 & 1.32 & 0.294 & 1.24 & 0.259 & 1.26 & 19.58 & 1.42 & 22.56 & 1.20 & 19.64 & 1.47 & 10.79 & 1.42 & 10.64 & 1.21 & 8.94 \\
cross-rsdp-256-fast & 1.21 & 0.258 & 1.21 & 0.272 & 1.38 & 0.288 & 1.25 & 11.62 & 1.50 & 14.45 & 1.39 & 13.73 & 1.23 & 6.40 & 1.46 & 7.83 & 1.20 & 6.72 \\
cross-rsdpg-128-balanced & 1.21 & 0.035 & 1.20 & 0.037 & 1.37 & 0.041 & 1.28 & 3.26 & 1.27 & 3.40 & 1.41 & 3.70 & 1.26 & 1.97 & 1.32 & 2.14 & 1.39 & 2.28 \\
cross-rsdpg-128-fast & 1.38 & 0.040 & 1.21 & 0.037 & 1.21 & 0.036 & 1.29 & 1.68 & 1.28 & 1.73 & 1.24 & 1.70 & 1.29 & 1.02 & 1.29 & 1.06 & 1.48 & 1.21 \\
cross-rsdpg-128-small & 1.38 & 0.040 & 0.91 & 0.028 & 1.21 & 0.036 & 1.42 & 12.93 & 1.28 & 12.20 & 1.27 & 12.07 & 1.17 & 6.44 & 1.26 & 7.32 & 1.26 & 7.26 \\
cross-rsdpg-192-balanced & 1.24 & 0.066 & 1.06 & 0.059 & 1.19 & 0.067 & 1.43 & 5.43 & 1.15 & 4.57 & 1.24 & 4.85 & 1.42 & 3.31 & 0.97 & 2.38 & 1.23 & 2.92 \\
cross-rsdpg-192-fast & 1.22 & 0.065 & 1.39 & 0.078 & 1.08 & 0.061 & 1.26 & 3.76 & 1.43 & 4.37 & 0.96 & 2.89 & 1.26 & 2.35 & 1.40 & 2.76 & 1.27 & 2.39 \\
cross-rsdpg-192-small & 1.24 & 0.066 & 1.22 & 0.068 & 1.35 & 0.077 & 1.30 & 18.19 & 1.27 & 18.70 & 1.15 & 16.13 & 1.27 & 11.13 & 1.44 & 13.17 & 1.09 & 9.75 \\
cross-rsdpg-256-balanced & 0.95 & 0.077 & 1.25 & 0.106 & 1.21 & 0.108 & 1.30 & 8.52 & 1.30 & 8.94 & 1.47 & 9.84 & 1.30 & 4.99 & 1.32 & 5.33 & 1.09 & 4.39 \\
cross-rsdpg-256-fast & 1.10 & 0.089 & 1.22 & 0.104 & 1.18 & 0.106 & 1.08 & 5.43 & 1.28 & 6.76 & 1.25 & 6.60 & 1.26 & 3.95 & 1.26 & 4.20 & 1.25 & 4.11 \\
\hdashline
ECDSA & 1.10 & 0.047 & 1.29 & 0.063 & 1.20 & 0.050 & 0.93 & 0.089 & 1.12 & 0.114 & 1.09 & 0.107 & 0.89 & 0.260 & 1.06 & 0.323 & 0.95 & 0.281 \\
RSA-2048 & 0.99 & 363 & 0.73 & 261 & 1.01 & 349 & 0.98 & 4.98 & 0.68 & 3.65 & 1.01 & 5.20 & 0.75 & 0.102 & 0.69 & 0.101 & 0.97 & 0.133 \\
\bottomrule
\end{tabular}
\end{table*}

%%% AUTO-GENERATED by paper/scripts/make_tables.py from
%%% results_power_2026-08-17_35-23.json -- DO NOT EDIT BY HAND.
%%% One row per algorithm, keygen/sign/verify as column-groups. $P$ (W) and $E$ (always mJ,
%%% field='e_marginal_per_iteration') are both baseline-subtracted (see inject_marginal_energy()).
%%% Native columns: value. Container/unikernel columns: value.
\begin{table*}[p]
\centering
\caption{Full KEM-primitive power and energy results (Structure as in Table~\ref{tab:full-energy-sig}).}
\label{tab:full-energy-kem}
\footnotesize
\setlength{\tabcolsep}{2pt}
\begin{tabular}{l | rrrrrr|rrrrrr|rrrrrr}
\toprule
 & \multicolumn{6}{c|}{Keygen} & \multicolumn{6}{c|}{Encaps} & \multicolumn{6}{c}{Decaps} \\
 & \multicolumn{2}{c}{Ntv.} & \multicolumn{2}{c}{Cnt.} & \multicolumn{2}{c}{Uk.} & \multicolumn{2}{c}{Ntv.} & \multicolumn{2}{c}{Cnt.} & \multicolumn{2}{c}{Uk.} & \multicolumn{2}{c}{Ntv.} & \multicolumn{2}{c}{Cnt.} & \multicolumn{2}{c}{Uk.} \\
Algorithm & P & E & P & E & P & E & P & E & P & E & P & E & P & E & P & E & P & E \\
\midrule
BIKE-L1 & 1.42 & 61.29 & 1.55 & 69.78 & 1.38 & 76.82 & 1.39 & 3.04 & 1.19 & 2.72 & 0.84 & 2.40 & 1.39 & 48.92 & 0.92 & 33.34 & 1.17 & 52.05 \\
BIKE-L3 & 1.46 & 195 & 1.41 & 199 & 1.38 & 239 & 1.39 & 9.40 & 1.10 & 7.55 & 1.28 & 11.25 & 1.39 & 156 & 1.57 & 181 & 1.00 & 140 \\
BIKE-L5 & 1.44 & 486 & 1.44 & 497 & 1.00 & 434 & 1.40 & 23.93 & 1.39 & 24.73 & 1.21 & 27.22 & 1.46 & 411 & 1.41 & 407 & 0.95 & 332 \\
Classic-McEliece-348864 & 1.23 & 397 & 1.55 & 548 & 1.67 & 562 & 1.31 & 0.220 & 1.35 & 0.242 & 1.14 & 0.297 & 1.05 & 60.91 & 1.07 & 64.67 & 0.97 & 54.76 \\
Classic-McEliece-348864f & 1.29 & 207 & 1.39 & 234 & 1.42 & 223 & 1.01 & 0.171 & 1.34 & 0.238 & 1.29 & 0.334 & 1.06 & 60.86 & 1.05 & 63.43 & 1.24 & 70.16 \\
Classic-McEliece-460896 & 1.64 & 2279 & 1.56 & 2096 & 1.33 & 1344 & 1.34 & 0.520 & 1.41 & 0.537 & 1.28 & 0.749 & 1.10 & 99.90 & 0.82 & 75.68 & 1.04 & 96.49 \\
Classic-McEliece-460896f & 1.45 & 772 & 1.48 & 805 & 1.51 & 842 & 1.32 & 0.474 & 1.54 & 0.580 & 1.28 & 0.756 & 1.06 & 92.27 & 1.32 & 121 & 1.04 & 96.49 \\
HQC-128 & 1.42 & 7.37 & 1.44 & 7.84 & 1.00 & 6.85 & 1.42 & 14.86 & 1.54 & 16.91 & 0.97 & 13.18 & 1.42 & 23.60 & 1.25 & 21.24 & 0.84 & 17.28 \\
HQC-192 & 1.44 & 22.99 & 1.45 & 23.94 & 1.29 & 27.43 & 1.45 & 46.18 & 1.41 & 47.22 & 1.23 & 51.37 & 1.44 & 70.29 & 1.41 & 72.25 & 1.01 & 63.03 \\
HQC-256 & 1.28 & 37.32 & 1.41 & 43.25 & 1.14 & 44.74 & 1.12 & 65.75 & 1.41 & 86.66 & 1.12 & 85.38 & 1.41 & 128 & 1.43 & 136 & 1.28 & 148 \\
Kyber512 & 1.43 & 0.067 & 1.00 & 0.049 & 1.24 & 0.108 & 1.17 & 0.061 & 1.29 & 0.072 & 1.26 & 0.132 & 1.17 & 0.050 & 1.32 & 0.060 & 1.27 & 0.151 \\
Kyber768 & 1.32 & 0.089 & 1.17 & 0.081 & 1.24 & 0.171 & 1.53 & 0.121 & 1.11 & 0.093 & 1.25 & 0.208 & 0.91 & 0.063 & 1.29 & 0.095 & 1.26 & 0.232 \\
Kyber1024 & 1.34 & 0.134 & 1.56 & 0.162 & 1.07 & 0.222 & 1.09 & 0.126 & 1.22 & 0.143 & 0.95 & 0.230 & 1.36 & 0.142 & 1.19 & 0.127 & 1.28 & 0.344 \\
ML-KEM-512 & 0.97 & 0.077 & 1.26 & 0.105 & 1.11 & 0.097 & 1.29 & 0.122 & 1.29 & 0.129 & 1.09 & 0.106 & 1.29 & 0.149 & 1.44 & 0.214 & 1.08 & 0.128 \\
ML-KEM-768 & 1.31 & 0.170 & 1.21 & 0.199 & 1.24 & 0.172 & 1.31 & 0.199 & 1.35 & 0.210 & 1.39 & 0.215 & 1.32 & 0.236 & 1.29 & 0.241 & 1.17 & 0.214 \\
ML-KEM-1024 & 1.12 & 0.223 & 1.28 & 0.264 & 1.27 & 0.265 & 1.10 & 0.244 & 1.27 & 0.297 & 1.26 & 0.289 & 1.21 & 0.314 & 1.30 & 0.357 & 1.30 & 0.347 \\
FrodoKEM-640-AES & 1.20 & 20.65 & 0.87 & 15.57 & 1.14 & 18.26 & 1.01 & 17.62 & 1.15 & 20.84 & 1.20 & 19.21 & 0.94 & 16.11 & 1.23 & 22.58 & 1.20 & 19.30 \\
FrodoKEM-640-SHAKE & 1.31 & 7.58 & 1.04 & 6.34 & 1.22 & 7.14 & 1.33 & 8.76 & 1.14 & 7.76 & 1.34 & 8.88 & 1.44 & 9.36 & 1.09 & 7.32 & 1.29 & 8.47 \\
FrodoKEM-976-AES & 1.16 & 45.71 & 1.15 & 47.56 & 1.05 & 38.33 & 1.20 & 48.08 & 1.32 & 55.13 & 1.33 & 49.07 & 1.19 & 47.22 & 0.99 & 40.04 & 1.19 & 43.97 \\
FrodoKEM-976-SHAKE & 1.28 & 16.78 & 1.30 & 17.93 & 1.45 & 18.90 & 1.32 & 19.32 & 1.32 & 20.09 & 1.13 & 16.41 & 1.35 & 20.17 & 1.31 & 19.73 & 1.08 & 15.71 \\
FrodoKEM-1344-AES & 1.00 & 74.23 & 1.17 & 91.27 & 1.18 & 81.16 & 1.23 & 93.10 & 1.21 & 96.17 & 1.22 & 85.65 & 1.20 & 91.28 & 1.21 & 95.38 & 1.36 & 95.38 \\
FrodoKEM-1344-SHAKE & 1.16 & 27.77 & 1.34 & 33.97 & 1.24 & 29.56 & 1.10 & 29.61 & 1.37 & 39.01 & 1.27 & 34.09 & 1.37 & 36.20 & 1.35 & 37.30 & 1.31 & 35.11 \\
\hdashline
ECDHE & 1.19 & 0.075 & 1.02 & 0.066 & 1.21 & 0.076 & 1.23 & 0.300 & 0.92 & 0.230 & 1.05 & 0.266 & --- & --- & --- & --- & --- & --- \\
X25519 & 1.22 & 0.102 & 1.24 & 0.108 & 1.27 & 0.095 & 1.02 & 0.261 & 1.00 & 0.260 & 1.00 & 0.244 & --- & --- & --- & --- & --- & --- \\
\hdashline[1pt/1pt]
x25519\_Kyber512 & 1.08 & 0.142 & 1.35 & 0.187 & 1.27 & 0.212 & 0.91 & 0.280 & 1.18 & 0.381 & 1.09 & 0.387 & 1.17 & 0.345 & 1.05 & 0.324 & 1.08 & 0.396 \\
x25519\_Kyber768 & 1.28 & 0.197 & 1.39 & 0.223 & 1.25 & 0.272 & 1.09 & 0.362 & 1.18 & 0.411 & 1.09 & 0.449 & 1.13 & 0.365 & 1.17 & 0.396 & 1.11 & 0.480 \\
x25519\_Kyber1024 & 1.28 & 0.238 & 1.29 & 0.249 & 1.13 & 0.322 & 1.09 & 0.404 & 1.22 & 0.469 & 1.26 & 0.615 & 1.09 & 0.394 & 0.94 & 0.348 & 1.14 & 0.584 \\
\bottomrule
\end{tabular}
\end{table*}

\section{Error Propagation and Statistical Uncertainty Analysis}
\paragraph{Type A: Statistical Evaluation of Uncertainty}

\begin{equation}
    u_A(x) = \frac{s(x)}{\sqrt{n}}
    \label{eq:u-a}
\end{equation}

\paragraph{Type B: Evaluation of Uncertainty Formulas}

\begin{equation}
	\label{eq:uncertainty_energy_of_energy}
	u_E(E)      = \sqrt{u_U^2(E) + u_I^2(E)}
\end{equation}

\begin{equation}
	\label{eq:uncertainty_time_of_energy}
	u_t(E)      = E \frac{a_t}{T_{op}\sqrt{3}}
\end{equation}

\begin{equation}
	\label{eq:uncertainty_jump}
	u_{jump}(E) = \sqrt{{(\frac{|\Delta P_s| \cdot \Delta t}{\sqrt{12}})}^2 + {(\frac{|\Delta P_e| \cdot \Delta t}{\sqrt{12}})}^2}
\end{equation}

\begin{equation}
	\label{eq:uncertainty_voltage_of_energy}
	u_U(E)      = \frac{\Delta t}{\sqrt{3}} \sum_i I_i (a_U U_i + b_U)
\end{equation}

\begin{equation}
	\label{eq:uncertainty_current_of_energy}
	u_I(E)      = \frac{\Delta t}{\sqrt{3}} \sum_i U_i (a_I I_i + b_I)
\end{equation}

\end{document}